\documentclass[aps, prl, twocolumn, longbibliography, superscriptaddress]{revtex4-2}

\usepackage{amsmath,amssymb,bm,graphicx}
\usepackage{dcolumn}
\usepackage{color}
\usepackage[utf8]{inputenc}
\usepackage[T1]{fontenc}
\usepackage[colorlinks=true,citecolor=blue]{hyperref}
\hypersetup{colorlinks=true,citecolor=blue,linkcolor=blue,urlcolor=blue}

\usepackage[english]{babel}
\usepackage[normalem]{ulem}
\usepackage{amsthm}
\usepackage{amsfonts}
\usepackage{mathtools}
\usepackage{enumerate}
\usepackage{color}
\usepackage[dvipsnames]{xcolor}
\usepackage{xifthen, xfrac}
\usepackage{cancel}
\usepackage{euscript} %, mathrsfs}
\usepackage{mathbbol} % for the x-notation of PSEs

\usepackage{dcolumn}% Align table columns on decimal point
\usepackage{bm}% bold math
\usepackage{siunitx}
\usepackage[textwidth=1.2cm, textsize=scriptsize]{todonotes}

\begin{document}

\title{Trion polaron problem in bulk and two-dimensional materials}

\author{V.~Shahnazaryan}
\email{vanikshahnazaryan@gmail.com}
\affiliation{Abrikosov Center for Theoretical Physics, MIPT, Dolgoprudnyi, Moscow Region 141701, Russia}

\author{A.~Kudlis}
\email{andrewkudlis@gmail.com}
\affiliation{Science Institute, University of Iceland, Dunhagi 3, IS-107, Reykjavik, Iceland}
\affiliation{Abrikosov Center for Theoretical Physics, MIPT, Dolgoprudnyi, Moscow Region 141701, Russia}

\author{K. Varga}
\email{kalman.varga@vanderbilt.edu}
\affiliation{Department of Physics and Astronomy, Vanderbilt University, Nashville, Tennessee, 37235, USA}

\author{I.~A.~Shelykh}
\affiliation{Science Institute, University of Iceland, Dunhagi 3, IS-107, Reykjavik, Iceland}

\author{I.~V.~Tokatly}
\affiliation{Donostia International Physics Center (DIPC), E-20018 Donostia-San Sebastián, Spain}
\affiliation{Nano-Bio Spectroscopy Group and European Theoretical Spectroscopy Facility (ETSF), Departamento de Polímeros y Materiales
Avanzados: Física, Química y Tecnología, Universidad del País Vasco, Avenida Tolosa 72, E-20018 San Sebastián, Spain}
\affiliation{IKERBASQUE, Basque Foundation for Science, 48009 Bilbao, Spain}

\begin{abstract}

We develop a microscopic theory of the \emph{trion–polaron}: a bound state of two electrons and one hole, dressed by longitudinal optical (LO) phonons. Starting from the Fr\"ohlich Hamiltonian, which describes the interaction of charged particles with LO phonons in three-dimensional (bulk) and two-dimensional (monolayer) polar crystals, we adopt the intermediate coupling variational approximation of Lee, Low, and Pines, and generalize it for the three-body problem. This yields an effective three‐particle Hamiltonian with renormalized electron–electron and electron–hole interactions, similar to those obtained for exciton polaron and bipolaron problems. We compute the binding energies for a family of bulk perovskite materials and several atomic monolayer materials characterized by pronounced polar effects, providing quantitative benchmarks for spectroscopic measurements.
\end{abstract}

\maketitle

Polarons--charge carriers dressed by a cloud of longitudinal optical (LO) phonons--are a cornerstone concept for transport and optical phenomena in polar crystals \cite{alexandrov2010advances}. 
The strong coupling (Landau–Pekar) model \cite{landau1933electron,pekar1946local,landau1948effective} and the field-theoretical Fr\"ohlich model with its celebrated Lee–Low–Pines (LLP) and Feynman treatments \cite{frohlich1950xx,lee1953motion,feynman1955slow}
constitute the fundamental basis underlying a vast body of theoretical and experimental studies in three-dimensional (3D) polar materials \cite{devreese2009frohlich,franchini2021polarons}. 
Recent methodological advances, including diagrammatic quantum Monte Carlo approach \cite{prokofev1998polaron,mishchenko2000diagrammatic,hahn2018diagrammatic}, and first-principles electron–phonon theory \cite{giustino2017electron,sio2019polarons,sio2019ab} allow for quantitative, materials-specific predictions.

In polar semiconductors multiparticle complexes of charge carriers  coupled to LO phonons, such as exciton polarons, are subject of intense studies by means of variational  \cite{haken1958theorie,mahanti1972effective,bajaj1974effect,pollmann1977effective,kane1978pollmann}, diagrammatic \cite{burovski2001diagrammatic,burovski2008exact,mishchenko2018optical,rana2024interplay}, and first-principal \cite{filip2021phonon,alvertis2024phonon,dai2024excitonic} approaches.
Metal–halide perovskites have recently emerged as a fertile platform where polaronic corrections shape optical spectra of excitons in both bulk and layered settings \cite{baranowski2020excitons,filip2021phonon,simbula2021polaron,tao2021momentarily,tao2022dynamic,dyksik2024polaron,biswas2024exciton,baranowski2024polaronic,lei2024persistent,duan20242d}. 

The rise of two-dimensional (2D) materials \cite{novoselov20162d} has brought qualitatively new physics. 
One of the key distinct features of 2D is the nonlocal screening of Coulomb forces.
For the Coulomb interactions of charge carriers this results in  Keldysh–Rytova potential \cite{Keldysh1979,Rytova1967,cudazzo2011dielectric}.
Another consequence of the Coulomb 2D screening is the nontrivial dispersion of LO phonons, predicted via first-principles calculations \cite{sohier2016two,sohier2017breakdown} and macroscopic electrostatic treatment \cite{shahnazaryan2025polarons}, and verified experimentally \cite{Li2024experiment}.
This peculiarity essentially reshapes the electron-phonon coupling in 2D, and has profound implications on polaron \cite{sio2022unified,sio2023polarons,kudlis2025polarons}, and exciton polaron \cite{alvertis2024phonon,lee2024phonon,shahnazaryan2025polarons} properties.

Beyond neutral exciton polarons, charged multiparticle polaron complexes are attracting growing interest. 
This particularly includes bipolarons \cite{alexandrov1981theory,adamowski1989formation,fomin1994bipolaron,macridin2004two,devreese2009frohlich}, which potentially can constitute an alternative mechanism of superconductivity \cite{alexandrov1981theory,zhang2023bipolaronic}.
While bipolarons remain elusive, stable three-body complexes, such as trions typically composed of two electrons and one hole, are well known in insulators and semiconductors.  In fact, under certain conditions trions dominate the optical spectra of 2D materials
\cite{mak2013tightly,wang2018colloquium} and hybrid layered perovskites \cite{ziegler2023mobile}. In polar materials, the interaction of LO phonons with trion states leading to the formation of trion polarons is naturally expected and has received considerable attention in recent years.
This includes experimental studies of trion polarons in perovskite quantum dots and nanoscrystals \cite{makarov2016spectral,fu2017neutral,becker2018bright,zhu2023many,tamarat2023universal,cho2024size}.
The key features of trion polarons in these studies are captured via variational treatment of three-body bound state problem \cite{movilla2025binding}, where the polaronic modulation of Coulomb interactions is accounted by means of phenomenological Bajaj-type potential \cite{bajaj1974effect}.

\begin{figure}
    \centering
    \includegraphics[width=0.99\linewidth]{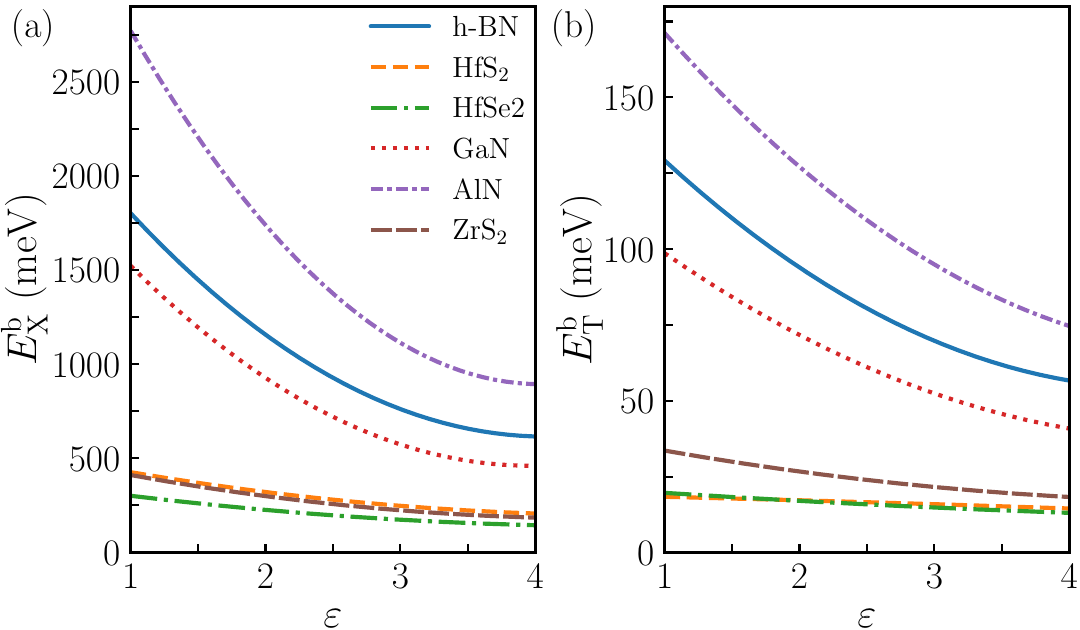}
    \caption{(a) Exciton polaron, and (b) trion polaron binding energy in polar atomic monolayers as a function of a dielectric constant of the surrounding media. }
    \label{fig:Eb-epsilon}
\end{figure}

\emph{In this work} we develop a systematic framework for the trion polaron problem. 
Building on the Lee–Low–Pines variational scheme \cite{lee1953motion} and the macroscopic description of LO phonons in 3D and 2D \cite{kittel1987quantum,sohier2017breakdown,shahnazaryan2025polarons}, we (i) derive an effective Hamiltonian for a charged exciton (two electrons and a hole) coupled to LO phonons, which encapsulates polaronic mass renormalization and phonon-modified, static effective electron–hole and electron–electron interactions; 
(ii) obtain closed analytical forms for the effective potentials in bulk, and integral representations tailored to nonlocal  screening in 2D; 
(iii) compute trion-polaron and exciton-polaron binding energies for representative lead halide perovskites and polar atomic monolayers using a stochastic variational solution of the internal trion problem, benchmarking against a Chandrasekhar-type ansatz; and (iv) map the dependence on the dielectric environment in 2D. 
Our results (Tables~\ref{table:3D_perovskites}–\ref{table:2D_materials} and Fig.~\ref{fig:Eb-epsilon}) reveal a robust hierarchy between bare, statically screened, and polaron-dressed complexes, a near-universal exciton to trion binding energy ratio in bulk perovskites, and remarkably large trion polaron binding in wide-gap polar monolayers.

%%%%%%%%%%%%%%%%%%%%%%%

\begin{table*}[t!]
\caption{Calculated statically screened ($E_{\rm T}^{\rm 0}$), trion polaron  ($E_{\rm T}^{\rm b}$), and bare ($E_{\rm T}^{\rm \infty}$) trion binding energies, statically screened ($E_{\rm X}^{\rm 0}$), exciton polaron  ($E_{\rm X}^{\rm b}$), and bare ($E_{\rm X}^{\rm \infty}$) exciton binding energies for several
lead trihalide perovskites, together with bare and renormalized
carrier masses ($m_{e,h}$/$m_{e,h}^*$), 
LO phonon energy ($\hbar\omega_{\rm l}$), 
static ($\epsilon_{\rm s}$) 
and high-frequency ($\epsilon_\infty$) dielectric constants, 
and Fr\"ohlich coupling constants $\alpha_{e,h}$.
Material parameters for Cs-based perovskites are from the Ref. \cite{filip2021phonon}, 
for MAPbI$_3$: Ref. \cite{baranowski2020excitons}, 
for MAPbBr$_3$: Ref. \cite{soufiani2015polaronic}.  }
\label{table:3D_perovskites}
\begin{ruledtabular}
\begin{tabular}{lccccccccc}
Material &
$E_{\rm T}^{0}$/$E_{\rm T}^{\rm b}$/$E_{\rm T}^{\infty}$ (meV) 
& $E_{\rm X}^{0}$/$E_{\rm X}^{\rm b}$/$E_{\rm X}^{\infty}$ (meV)  
&$m_e/m_e^*$ & $m_h/m_h^*$ &
$\hbar\omega_{\rm l}$ (meV) &
$\epsilon_{\rm s}$ & $\epsilon_\infty$ &
$\alpha_e$ & $\alpha_h$ \\ [3pt]\hline
MAPbCl$_3$ & 0.142/{\bf 6.33}/7.10 & 2.95/{\bf 84.3}/147 & 0.39\cite{bokdam2016role}/0.63 & 0.38\cite{bokdam2016role}/0.62 & 20.0\cite{nagai2018longitudinal} & 29.8\cite{sendner2016optical}  & 4.22\cite{bokdam2016role} & 3.31 & 3.27 \\
MAPbBr$_3$ & 0.138/{\bf 1.95}/4.64 & 2.96/{\bf 35.2}/99.4 & 0.26/0.36 & 0.31/0.45 & 22.0 & 25.5 & 4.40 & 2.38 & 2.60 \\
MAPbI$_3$  & 0.164/{\bf 0.90}/2.62 & 3.50/{\bf 18.0}/56.1 & 0.19/0.25 & 0.23/0.30 & 16.5 & 20.0 & 5.00 & 1.88 & 2.04 \\
CsPbCl$_3$ & 0.298/{\bf 3.77}/6.66 & 6.31/{\bf 66.4}/141 & 0.27/0.38 & 0.30/0.43 & 26.0 & 17.5 & 3.70 & 2.53 & 2.67 \\
CsPbBr$_3$ & 0.235/{\bf 2.05}/4.02 & 5.00/{\bf 37.2}/85.4 & 0.24/0.33 & 0.27/0.38 & 18.0 & 18.6 & 4.50 & 2.27 & 2.41 \\
CsPbI$_3$  & 0.143/{\bf 0.91}/2.40 & 3.07/{\bf 17.5}/51.3 & 0.21/0.28 & 0.25/0.34 & 14.0 & 22.5 & 5.50 & 1.96 & 2.14 \\
\end{tabular}
\end{ruledtabular}
\end{table*}
\begin{table}
\caption{Exciton ($E_{\rm X}^{\rm b}$) and trion ($E_{\rm T}^{\rm b}$) binding energies obtained with the SVM and Chandrasekhar ansatz (Ch) for several
lead trihalide perovskites.}
\label{table:Chandrasekhar}
\begin{ruledtabular}
\begin{tabular}{lcccc}
Material &
$E_{\rm X}^{\rm b}$ & $E_{\rm X}^{\rm Ch}$ &
$E_{\rm T}^{\rm b}$ & $E_{\rm T}^{\rm Ch}$   \\
 & (meV) & (meV) & (meV) & (meV) \\[3pt]\hline
MAPbCl$_3$ & 84.3 & 83.9 & 6.33 & 2.27  \\
MAPbBr$_3$ & 35.2 & 34.5 & 1.95 & 0.44 \\
MAPbI$_3$  & 18.0 & 17.4 & 0.90 & 0.26 \\
CsPbCl$_3$ & 66.4 & 65.7 & 3.77 & 1.15  \\
CsPbBr$_3$ & 37.2 & 36.8 & 2.05 & 0.62 \\
CsPbI$_3$  & 17.5 & 17.0 & 0.91 & 0.25 \\
\end{tabular}
\end{ruledtabular}
\end{table}
%

%%%%%%%%%%%%%%%%%%%%%%%
\begin{table*}[t!]
\caption{Calculated statically screened ($E_{\rm T}^{\rm 0}$), trion polaron  ($E_{\rm T}^{\rm b}$), and bare ($E_{\rm T}^{\rm \infty}$) trion binding energies, statically screened ($E_{\rm X}^{\rm 0}$), exciton polaron  ($E_{\rm X}^{\rm b}$), and bare ($E_{\rm X}^{\rm \infty}$) exciton binding energies  for several 2D polar materials. 
Material parameters including TO phonon energy ($\hbar \omega_{\rm t}$), static ($r_0$) and high-frequency ($r_\infty$) screening lengths, bare effective masses ($m_{e,h}$) are taken from the first principle calculations and experimental data (from Ref. \cite{sio2023polarons} if no other source is specified). 
The screening lengths are extracted from Ref. \cite{sio2023polarons} as
$r_{0[\infty]} = \epsilon_{0[\infty]} d/2$, 
where $\epsilon_{0[\infty]}$ is the static [high-frequency] dielectric constant, $d$ is the monolayer thickness.
The exciton binding energies and renormalized effective masses are from the Ref. \cite{shahnazaryan2025polarons}.
 }
\label{table:2D_materials}
\begin{ruledtabular}
\begin{tabular}{lccccccc}
Material & $E_{\rm T}^0$/$E_{\rm T}^{\rm b}$/$E_{\rm T}^\infty$ (meV) 
& $E_{\rm X}^0$/$E_{\rm X}^{\rm b}$/$E_{\rm X}^\infty$ (meV) 
& $m_e/m_e^*$ & $m_h/m_h^*$ & $\hbar\omega_{\rm t}$ (meV) & $r_0$ (nm) & $r_\infty$ (nm) \\[3pt]\hline
hBN     & $110.2/{\bf 130.2}/141.9$ & $1617/{\bf 1865}/2007$  & $0.83/1.00$ & $0.65/0.77$ & 172.3\cite{sohier2017breakdown}                        & 1.076\cite{sohier2017breakdown} & 0.780\cite{sohier2017breakdown} \\
GaN      & $100.3/{\bf 98.14}/132.0$ & $1304/{\bf 1586}/1660$   & $0.24/0.29$ & $1.35/2.02$ &  73.30\cite{Sanders2017}                               & 1.109 & 0.756 \\
AlN      & $149.3/{\bf 170.8}/215.0$ & $2064/{\bf 2873}/2839$  & $0.51/0.71$ & $1.49/2.53$ &  74.15\cite{doi:10.1021/acs.jpcc.6b09706}             & 0.763 & 0.466 \\
HfSe$_2$ & $7.138/{\bf 18.25}/28.29$ & $127.5/{\bf 306.9}/421.4$  & $0.18/0.43$ & $0.23/0.57$ &  11.10\cite{Li_2024}                                   & 21.75 & 4.246 \\
HfS$_2$  & $11.27/{\bf 16.83}/41.48$ & $197.8/{\bf 434.2}/615.6$  & $0.24/0.54$ & $0.44/1.07$ &  17.81\cite{doi:10.1021/acsomega.1c04286}            & 13.98 & 2.974 \\
ZrS$_2$  & $13.88/{\bf 32.26}/42.01$ & $238.5/{\bf 420.6}/619.2$  & $0.31/0.58$ & $0.26/0.48$ &  29.82\cite{Pandit_2021}                               & 10.62 & 2.826 \\
\end{tabular}
\end{ruledtabular}
\end{table*}
%

%%%%%%%%%%%%%%%%%%%%%%%

\textbf{The model}.
We consider the bound state problem for the system  consisting of two identical conduction electrons of mass $m_{e}$ and a valence band hole of mass $m_{h}$ subjected to a polar crystal. 
Along with the interparticle Coulomb interactions, the charge carriers interact with the electric field generated by LO phonons.
The respective Hamiltonian of resulting trion polaron state is
\begin{align}
\hat{H}
  &= \sum_{i=1}^{2}\left( \frac{\hat{\mathbf{p}}_{e i}^{2}}{2m_{e}}
  -V_{\rm C} \left(|\mathbf{r}_{e i}-\mathbf{r}_{h}|\right) \right) \notag \\
  &+ \frac{\hat{\mathbf{p}}_{h}^{2}}{2m_{h}}
  + V_{\rm C}\left(|\mathbf{r}_{e1}-\mathbf{r}_{e2}|\right)
  +\sum_{\mathbf{k}}
   \hbar\omega_{{\rm l},k}\,
    \hat{a}_{\mathbf{k}}^{\dagger}\hat{a}_{\mathbf{k}} \notag\\
  &+\sum_{\mathbf{k}}V_{k} \left( 
   \Bigl[
        e^{i\mathbf{k} \cdot \mathbf{r}_{e1}}
       +e^{i\mathbf{k} \cdot \mathbf{r}_{e2}}
       -e^{i\mathbf{k} \cdot \mathbf{r}_{h}}
   \Bigr]\hat{a}_{\mathbf{k}}
   +\text{h.c.} \right) .
\label{eq:H_start}
\end{align}
$m_{e[h]}$, $\hat{\mathbf{p}}_{e[h]}$, $\mathbf{r}_{e[h]}$ are the electron [hole] mass, momentum, and position, respectively,
$\hat{a}_{\mathbf{k}}^{\dagger}$ creates a phonon with momentum~$\mathbf{k}$ and frequency $\omega_{{\rm l},k}$,
$V_{\rm C}$ is the Coulomb potential accounting for the \textit{electronic} screening only, and $V_k$ is the electron-phonon coupling, both specified later.
We handle the problem variationally in the virtue of single polaron LLP transformation \cite{lee1953motion}.
The LLP approach was previously generalized to describe the bulk \cite{pollmann1977effective,kane1978pollmann} and 2D \cite{shahnazaryan2025polarons} exciton polaron problem. 
Here we extend the treatment for the charged exciton complex.
We proceed with the relative 
\begin{align}
\bm{\rho}_{1}= \mathbf{r}_{e1}-\mathbf{r}_{h},\quad
\bm{\rho}_{2}= \mathbf{r}_{e2}-\mathbf{r}_{h},
\end{align}
and center–of–mass (COM)
\begin{align}
\mathbf{R}=\beta_{e}\mathbf{r}_{e1}
          +\beta_{e}\mathbf{r}_{e2}
          +\beta_{h}\mathbf{r}_{h}
\end{align}
coordinates, where $\beta_{e[h]}=m_{e[h]}/M$, and
$M = 2m_e +m_h$.
Applying a unitary transformation
$\hat{S}= \exp \Bigl\{
 i \bigl[\mathbf{Q}
   -\sum_{\mathbf{k}} \mathbf{k}
\hat{a}_{\mathbf{k}}^{\dagger} \hat{a}_{\mathbf{k}}\bigr]
  \cdot \mathbf{R} \Bigr\}$
one gets rid of COM coordinate $\mathbf{R}$, yielding a {\it C}-number for the COM momentum $\hbar \mathbf{Q}$. 
Next, we apply a phonon field shift unitary operator 
$\hat{U}=\exp
\Bigl[
  \sum_{\mathbf{k}}
   \bigl(
     \hat{a}_{\mathbf{k}}F_{\mathbf{k}}^{*}
    -\hat{a}_{\mathbf{k}}^{\dagger}F_{\mathbf{k}}
   \bigr)
\Bigr]$,
acting as $\hat{U}^{-1}\hat{a}_{\mathbf{k}}\hat{U}
 =\hat{a}_{\mathbf{k}}+F_{\mathbf{k}}$.
Here $F_{\mathbf{k}} = F_{\mathbf{k}} (\bm{\rho}_{1}, \bm{\rho}_{2})$ is a variational parameter corresponding to the mean displacement. 
We seek the eigenfunction as the ground-state for phonon annihilation operator: $\hat{a}_{\mathbf{k}} |\Psi \rangle =0$.
Performing a variational minimization with respect to $F_{\mathbf{k}}$, we obtain the following effective Hamiltonian for the trion polaron internal dynamics [see {\color{blue}Supplemental Material (SM)} for the derivation]:
\begin{align}
\hat{H}_{\rm eff}
  &=  \frac{\hat{\mathbf{p}}_{1}^{2}}{2\mu^*}
    +\frac{\hat{\mathbf{p}}_{2}^{2}}{2\mu^*}
    +\frac{2\sigma}{\sigma+1}
\frac{\hat{\mathbf{p}}_{1}\!\cdot\!\hat{\mathbf{p}}_{2}}{2\mu^*} \notag\\
    &+V_{eh}^{\rm eff}(\rho_{1})
    +V_{eh}^{\rm eff}(\rho_{2})
    +V_{ee}^{\rm eff}(|\bm{\rho}_{1}-\bm{\rho}_{2}|).
\label{eq:H_eff_final}
\end{align}
Here the many–body effects of the polar lattice are encoded both in the reduced mass $\mu^* = m_e^* m_h^* / (m_e^* + m_h^*)$ accounting for the polaron renormalization of electron and hole masses,
and in the static potentials 
$V_{eh}^{\rm eff}$ and $V_{ee}^{\rm eff}$,
which read as
\begin{align}
    \label{eq:Veff0eh}
    &V_{eh}^{\rm eff } (\rho) = -V_{\rm C} (\rho) \notag \\
    &+\sum_k \cos( \mathbf{k} \cdot \bm{\rho}) 
     \frac{2 |V_k|^2}{\Delta m}
    \left( \frac{m_h}{  \hbar \omega_{{\rm l}, k}  + \frac{\hbar^2 k^2}{2m_h} }
    - \frac{m_e}{  \hbar \omega_{{\rm l}, k}  + \frac{\hbar^2 k^2}{2m_e} }\right) ,  \\
    \label{eq:Veff0ee}
    &V_{ee}^{\rm eff} (\rho) = V_{\rm C} (\rho) 
    -\sum_k 2 |V_k|^2 \cos( \mathbf{k} \cdot \bm{\rho}) 
    \frac{  \hbar \omega_{{\rm l}, k} +2 \frac{\hbar^2 k^2}{2m_e} }
    {\left(\hbar \omega_{{\rm l}, k} +\frac{\hbar^2 k^2}{2m_e}\right)^2} ,
\end{align}
where $\Delta m = m_h - m_e$. 
As expected,
$ V_{eh}^{\rm eff } (\rho) \rightarrow  - V_{ee}^{\rm eff } (\rho)$ in the limit of $m_h \rightarrow m_e$
We further proceed with the explicit expressions for the effective Coulomb potentials in 3D and 2D crystals.

\textit{Bulk media} -- In bulk crystals the bare Coulomb potential is simply $V_{\rm C}^{\rm 3D}(\rho) = e^2 / (4\pi\varepsilon_0 \epsilon_\infty)$, where $\varepsilon_0$ is the vacuum permittivity,
and $\epsilon_\infty$ is the high-frequency dielectric constant.
Here optical phonons in the relevant wave vector range possess trivial dispersion, $\omega_{{\rm l}, k}  \approx \omega_{\rm l}$. 
The electron-phonon coupling is 
$V_k^{\rm 3D} = -i / k  \sqrt{ e^2 \hbar \omega_{\rm l} / (2\varepsilon_0 \epsilon^* V) } $,
where $1/\epsilon^* = 1/\epsilon_\infty - 1/\epsilon_{\rm s}$, $\epsilon_{\rm s}$ is the static dielectric constant, and $V$ is the sample volume.
The summation in Eqs.~\eqref{eq:Veff0eh}, \eqref{eq:Veff0ee} yields in
\begin{align}
\label{eq:Veh_bulk}
V_{eh}^{\rm eff,3D}(\rho)
 &= -\frac{e^{2}}{4\pi\varepsilon_{0}\rho}
    \Bigl[
      \frac{1}{\epsilon_{\rm s}}
     + \frac{1}{\epsilon^{*}}
        \Bigl(
           \frac{m_{h}}{\Delta m}\,e^{-\rho/l_{h}}
          -\frac{m_{e}}{\Delta m}\,e^{-\rho/l_{e}}
        \Bigr)
    \Bigr],
\\
\label{eq:Vee_bulk}
V_{ee}^{\rm eff,3D}(\rho)
 &=  \frac{e^{2}}{4\pi\varepsilon_{0}\rho}
    \Bigl[
      \frac{1}{\epsilon_{\rm s}}
     +\frac{1}{\epsilon^{*}}
      \Bigl( 1-\frac{\rho}{2l_{e}}\Bigr)
       e^{-\rho/l_{e}}
    \Bigr],
\end{align}
where $l_{e[h]} = \sqrt{ \hbar / (2\omega_{\rm l} m_{e[h]} }$ is the polaron radius. The electron-hole interaction of Eq~\eqref{eq:Veh_bulk} is exactly that of the exciton polaron problem \cite{pollmann1977effective}. The electron-electron interaction Eq.~\eqref{eq:Vee_bulk} coincides with the electron-hole interaction in the limit of identical effective masses.
On the other hand, it corresponds to the weak coupling limit of the effective interaction in the variational treatment of bipolaron problem  \cite{adamowski1989formation}.

\textit{2D media}-- 
In atomic semiconducting monolayers the bare Coulomb potential neglecting the lattice effects is given by Keldysh-Rytova potential \cite{Rytova1967,Keldysh1979,cudazzo2011dielectric}
\begin{align}
    V_{\rm C}^{\rm 2D} (\rho,r_\infty) = \frac{e^2}{4\pi\varepsilon_0} \frac{\pi}{2 r_\infty} 
    \left[ H_0\left( \frac{\varepsilon \rho}{r_\infty} \right) - Y_0\left( \frac{\varepsilon \rho}{r_\infty} \right) \right] ,
\end{align}
where $r_\infty$ is the high-frequency screening length, $\varepsilon$ is the dielectric constant of surrounding media, $H_0$, $Y_0$ are the Struve function and Bessel function of the second kind, respectively.
The electron-phonon coupling here is \cite{shahnazaryan2025polarons}
\begin{align}
    V_k^{\rm 2D} = -i e  \omega_{\rm t} 
    \sqrt{ \frac{\hbar (r_0 - r_\infty) }{4 \varepsilon_0 A \omega_{{\rm l}, k} } } \frac{ 1}{\varepsilon +  r_\infty k },  
\end{align}
where $r_0$ is the static screening length, $A$ is the sample area, $\omega_{{\rm l}, k} = \omega_{{\rm t}} \sqrt{(\varepsilon + r_0 k)/ (\varepsilon + r_\infty k)}$ is the LO phonon dispersion with $\omega_{\rm t}$ denoting the transverse optical (TO) phonon frequency \cite{sohier2017breakdown,shahnazaryan2025polarons}.
The complex dispersion relation for LO phonons doesn't allow to get analytical expressions for the effective Coulomb potentials, which are written in the following form:
\begin{align}
    \label{eq:Veh2D}
     &V_{eh}^{\rm eff, 2D} (\rho) =    
     -\frac{e^2 }{ 4 \pi \varepsilon_0 r_\infty}
    \int\limits_0^\infty   \frac{J_0 ( x \varepsilon \rho/r_\infty)}{1 + x } \times \notag \\
    & \left[ \frac{1}{x} - \frac{ \sigma_0-1 }{ (1 +  \sigma_0 x ) {\Delta m}  }  
      \left( \frac{m_h}{ f_{\sigma_{h,{\rm t}}} (x) }
    - \frac{m_e}{ f_{\sigma_{e,{\rm t}}} (x) }\right) \right]  x{\rm d} x , \\
    \label{eq:Vee2D}
    &V_{ee}^{\rm eff, 2D} (\rho) 
    =  \frac{e^2 }{ 4 \pi \varepsilon_0 r_\infty}
    \int\limits_0^\infty  \frac{J_0 (x \varepsilon \rho/r_\infty)}{1 + x } \times \notag \\
    &\left[ \frac{1}{x}-  \frac{ \sigma_0-1 }{ 1 +  \sigma_0 x   }  
    \frac{2 f_{\sigma_{e,{\rm t}}} (x) -1 }
    { f_{\sigma_{e,{\rm t}}}^2 (x) } \right]  x{\rm d} x ,
\end{align}
where $\sigma_0 = r_0 / r_\infty$, 
$\sigma_{i,{\rm t}} = \varepsilon r_{i,{\rm t}}/ r_\infty$,
$r_{i,{\rm t}} = \sqrt{\hbar/(2m_i\omega_{\rm t})}$,
$f_{\sigma_{i,{\rm t}}} (x) = 1  + \sigma_{i,{\rm t}}^2 x^2  \sqrt{(1+x)/(1+\sigma_0 x)}$, $i=e,h$.
Substituting  $\sqrt{(1+x)/(1+\sigma_0 x)} \approx 1 /\sqrt{\sigma_0}$ one can get approximate expressions for interaction potentials, as discussed in {\color{blue} SM}.

\textit{Numerical techniques}--
We tackle the effective Hamiltonian of Eq.~\eqref{eq:H_eff_final} numerically,
invoking the stochastic variational method (SVM)~\cite{SuzukiVarga1998}.
We seek the ground state as a linear combination of basis functions in a form of fully correlated Gaussians
\begin{align}\label{eq:psi_svm}
\psi_{\rm SVM}=& \sum_{n=1}^{N}
   c_{n} \Phi_n\nonumber\\
 =& \sum_{n=1}^{N}
   c_{n}\,
   \exp\!\left[
     -\tfrac{1}{2}
      \sum_{\alpha,\beta=1}^{2}
       A_{n}^{\alpha\beta}\,
       \bm{\rho}_{\alpha}\!\cdot\!\bm{\rho}_{\beta}
   \right]\chi_{SM_S},
\end{align}
where the positive–definite matrices
$A_{n}$ are generated and refined stochastically,
$\chi_{SM_S}$ is the coupled spin function, and the $c_{n}$’s are
linear variational coefficients~\cite{SuzukiVarga1998}. 
The effective Hamiltonian~\eqref{eq:H_eff_final} can be viewed as a system of two identical particles, which implies the opposite spin orientations for the ground state solution.
This choice renders the spatial part of Eq.~\eqref{eq:psi_svm} symmetric.
%, so the antisymmetrization need not be written explicitly.
Since the Hamiltonian contains a cross-kinetic term that couples the two identical particles, the kinetic-energy matrix acquires off-diagonal elements already at the operator level. 
All such terms are evaluated analytically within the Gaussian basis, following the recipes of Ref.~\cite{SuzukiVarga1998}.

The variational optimization consists in an iterative random modification of
$A_{n}^{\alpha\beta}$ until the ground-state energy converges. 
Since the basis functions are non-mutually orthogonal, we solve a generalized eigenvalue problem 
\begin{align}
\sum_{j}(H_{ij}-E\,O_{ij})\,c_{j}=0,\label{eq:egvprob_gen}
\end{align}
where matrix elements are defined as $H_{ij}=\langle\Phi_{i}|\hat{H}|\Phi_{j}\rangle$ and $O_{ij}=\langle\Phi_{i}|\Phi_{j}\rangle$. The solution of~\eqref{eq:egvprob_gen} yields the variational upper bounds $E_{0},E_{1},\dots$, and the
corresponding eigenvectors. %Finally, a few words on how the contribution of interaction matrix elements in $H_{ij}$ were evaluated.
Finally, we present the evaluation scheme for the contribution of interaction matrix elements in $H_{ij}$.
For the three-dimensional potentials of Eqs.
\eqref{eq:Veh_bulk},
\eqref{eq:Vee_bulk}– the required integrals can be reduced to closed analytic expressions in terms of standard special functions, and we adopt those results directly in the code. The two-dimensional potentials of Eqs.~\eqref{eq:Veh2D}, \eqref{eq:Vee2D} do not allow such simplifications; here we evaluate the integrals numerically with high-order Gauss–Legendre quadrature, which provides rapid and reliable convergence.

%%%%%%%%%%%%%%%%%%%%%%%
\textbf{Results}.
We proceed with material-specific calculations in the domains of bulk and 2D media.

\textit{Bulk media}.
Provided by the recent topmost research interest, we
focus on perovskites as representative bulk polar materials. 
Namely, we consider lead trihalide materials, both in all-inorganic (Cs-based) and hybrid (Methylammonium, MA) configurations. 
The trion polaron binding energies are presented in Table~\ref{table:3D_perovskites}.
For the reference we show the exciton binding energies, calculated on equal footing accounting for the effective mass polaronic renormalization, and the electron-hole interaction represented by Pollmann-Buttner potential \eqref{eq:Veh_bulk}.
One can see a universal ratio $E_{\rm X}^{\rm b}/ E_{\rm T}^{\rm b} \approx 20$ for the family of lead trihalide perovskites, which is similar to the respective scaling for transition metal dichalcogenide monolayers \cite{courtade2017charged}.
The little deviation from this scaling for MAPbCl$_3$ is tentatively due to the nearly identical electron and hole effective masses.
We also calculate the trion and exciton binding energies using Coulomb potentials accounting for the bare electron screening $\varepsilon_\infty$, and the static screening $\varepsilon_{\rm s}$, as shown in Table~\ref{table:3D_perovskites}.
As expected, both for exciton and trion the polaronic effects result in the binding energies lying in between these two limiting estimates.

To get further insight, we employ a simplistic Chandrasekhar type variational anzats for trion polaron wave function, and a respective hydrogen-like anzats for exciton polaron wave function (see the {\color{blue} SM}).
The resulting values for binding energy are shown in Table~\ref{table:Chandrasekhar}. 
While for exciton the hydrogen anzats gives a good match with the SVM calculation, the trion polaron binding energy results are strongly underestimated as compared to accurate SVM calculation.

\textit{2D media}.
We consider the trion polaron states in a set of atomically thin materials characterized by significant electron-phonon coupling, where previously the single-polaron \cite{sio2023polarons,shahnazaryan2025polarons} and exciton polaron \cite{shahnazaryan2025polarons} states were studied.
To treat all the materials on an equal footing, we set the dielectric constant of surrounding media $\varepsilon=1$.
The values of trion polaron binding energy are shown in Table~\ref{table:2D_materials}.
One can see again the order of magnitude difference as compared to exciton polaron binding energy for all the materials. 
The trion polaron and exciton polaron binding energies appear in between the limiting cases of high-frequency and static dielectric screening, similar to the bulk materials.
Surprisingly, for GaN this is not the case: the trion polaron binding is lower than the trion binding energy with  static screening. 
The reason for this is in large contrast of electron and hole effective masses, resulting in substantially different screening of the electron-hole and electron-electron interactions.
Particularly, as shown in {\color{blue} SM} Fig. S1, at small interparticle distances the electron-electron repulsion is stronger, as compared to electron-hole attraction, which results in the reduction of trion polaron binding energy.

We also note, that the polaron mass renormalization additionally enhances the binding energies of exciton and trion polarons.
This particularly for AlN results in larger binding energy of exciton polaron as compared to the exciton with high-frequency (electronic) dielectric screening. 
We therefore  for clarity perform the calculation of trion polaron and exciton polaron binding energies discarding the effective mass renormalization, as shown in {\color{blue} SM} Table~SI.
The results indicate no anomalous behavior in the hierarchy of exciton and trion binding energies apart from the AlN monolayer.

%%%%%%%%%%%%%%%%%%%%%%%
\textbf{Discussion and Conclusion}.
The obtained values of trion polaron binding energies for lead trihalide perovskites are of the 1-6 meV range.
At the same time, the exciton polaron peaks here are characterized by tens of meV linewidth \cite{soufiani2015polaronic,shi2020exciton}.
This fact essentially complicates the experimental observation of trion polaron states in bulk perovskites, requiring a careful multi-Lorentz fit of the peaks corresponding to exciton complexes with the resolution below the linewidth.
On the contrary, trion peaks are well visible in perovskite nanocrystals and quantum dots (QD) \cite{makarov2016spectral,fu2017neutral,becker2018bright,zhu2023many,tamarat2023universal,cho2024size}.
Here the size quantization of each individual charge carrier adds several meV to binding energy, thus providing an additional splitting between trion and exciton states. 
For the CsPbBr$_3$ QDs in the wide lateral size limit (about 30 nm) the trion binding energy is about 7 meV \cite{zhu2023many}.
This value is comparable with our result, and  can be reproduced within the trion polaron approach accounting for the size quantization effect \cite{movilla2025binding}.

In the 2D domain for the wide bandgap materials the obtained trion polaron binding energies are above 100 meV. 
One has to keep in mind, however, that the results are for freestanding monolayers, which imply the presence of undesired bending deformation. 
One therefore has to consider the presence of substrate with larger bandgap to secure that the free carriers are confined within the monolayer.
While this will introduce additional dielectric screening of the Coulomb interactions together with the possible presence of surface optical phonon mode, 
the obtained very large value of trion binding energy facilitates their possible observation.
This is confirmed by the exciton polaron and trion polaron binding energy dependence on a surround media's dielectric constant, shown, in Fig.~\ref{fig:Eb-epsilon}.
The most common material meeting all the above listed criteria is Si/SiO$_2$, which is widely used as a substrate for 2D materials \cite{chernikov2014exciton}.
Other options include  the use of LiF substrate, provided that a stable heterostructure hBN/LiF was previously obtained \cite{xie2017stitching}.
Another perspective direction is the utilization of twisted hBN homostructures, where recently promising results for exciton photoluminescence were obtained \cite{roux2025exciton}.

\textit{In conclusion}, we have formulated a microscopic theory of \emph{trion polarons} in both bulk and atomically thin polar materials. Starting from a macroscopic description of LO modes and the associated Fr\"ohlich coupling \cite{kittel1987quantum,sohier2017breakdown,shahnazaryan2025polarons}, we extended the Lee–Low–Pines variational technique \cite{lee1953motion} to a three-body charged complex. 
The resulting effective Hamiltonian combines polaronic mass renormalization with phonon-induced, static electron–hole and electron–electron interactions.
In 3D, the interactions acquire Yukawa-like analytic forms determined by polaron radii, while in 2D they are expressed through screened integral kernels consistent with nonlocal screening \cite{Keldysh1979,Rytova1967,cudazzo2011dielectric}. 

Numerical solutions of the effective trion problem, performed with a stochastic variational method and benchmarked against a Chandrasekhar ansatz, yield material trends across lead halide perovskites and polar monolayers. In bulk perovskites, trion-polaron binding energies fall in the 1–6~meV range and track a nearly constant $E_{\rm X}^{\rm b}/E_{\rm T}^{\rm b}\!\approx\!20$, consistent with the markedly stronger binding of neutral exciton polarons. 
In 2D wide-gap polar monolayers the trion-polaron binding is enhanced by reduced dimensionality and nonlocal screening, remaining large even when the dielectric environment is accounted for. We also clarified regimes where polaronic mass renormalization and phonon-induced electron–electron repulsion conspire to nontrivially reshape the trion spectrum. 

Beyond establishing a compact and versatile framework for charged exciton–phonon physics, our results provide practical guidance for experiment. In bulk perovskites, the small energy scales imply that resolving trion-polaron lines demands spectral analyses beyond the typical exciton linewidths, whereas in polar monolayers the predicted binding energies make trion polarons promising targets for optical spectroscopy and many-body engineering. 
We anticipate that this framework will enable quantitative, materials-specific studies of charged exciton complexes in both 3D perovskites and 2D polar crystals. 

\textit{Acknowledgements}. 
The research is supported by the
Ministry of Science and Higher Education of the Russian
Federation (Goszadaniye) Project No. FSMG-2023-0011.
V.S. acknowledges the support of “Basis” Foundation (Project No. 25-1-3-11-1). 
The work of A.K. is supported by the Icelandic Research
Fund (Rannsóknasjóður, Grant No. 2410550). IVT
acknowledges support from the Spanish MCIN/AEI/
10.13039/501100011033 through the project PID2023-
148225NB-C32, and the Basque Government (Grant No.
IT1453-22).


\begin{thebibliography}{79}%
\makeatletter
\providecommand \@ifxundefined [1]{%
 \@ifx{#1\undefined}
}%
\providecommand \@ifnum [1]{%
 \ifnum #1\expandafter \@firstoftwo
 \else \expandafter \@secondoftwo
 \fi
}%
\providecommand \@ifx [1]{%
 \ifx #1\expandafter \@firstoftwo
 \else \expandafter \@secondoftwo
 \fi
}%
\providecommand \natexlab [1]{#1}%
\providecommand \enquote  [1]{``#1''}%
\providecommand \bibnamefont  [1]{#1}%
\providecommand \bibfnamefont [1]{#1}%
\providecommand \citenamefont [1]{#1}%
\providecommand \href@noop [0]{\@secondoftwo}%
\providecommand \href [0]{\begingroup \@sanitize@url \@href}%
\providecommand \@href[1]{\@@startlink{#1}\@@href}%
\providecommand \@@href[1]{\endgroup#1\@@endlink}%
\providecommand \@sanitize@url [0]{\catcode `\\12\catcode `\$12\catcode `\&12\catcode `\#12\catcode `\^12\catcode `\_12\catcode `\%12\relax}%
\providecommand \@@startlink[1]{}%
\providecommand \@@endlink[0]{}%
\providecommand \url  [0]{\begingroup\@sanitize@url \@url }%
\providecommand \@url [1]{\endgroup\@href {#1}{\urlprefix }}%
\providecommand \urlprefix  [0]{URL }%
\providecommand \Eprint [0]{\href }%
\providecommand \doibase [0]{https://doi.org/}%
\providecommand \selectlanguage [0]{\@gobble}%
\providecommand \bibinfo  [0]{\@secondoftwo}%
\providecommand \bibfield  [0]{\@secondoftwo}%
\providecommand \translation [1]{[#1]}%
\providecommand \BibitemOpen [0]{}%
\providecommand \bibitemStop [0]{}%
\providecommand \bibitemNoStop [0]{.\EOS\space}%
\providecommand \EOS [0]{\spacefactor3000\relax}%
\providecommand \BibitemShut  [1]{\csname bibitem#1\endcsname}%
\let\auto@bib@innerbib\@empty
%</preamble>
\bibitem [{\citenamefont {Alexandrov}\ and\ \citenamefont {Devreese}(2010)}]{alexandrov2010advances}%
  \BibitemOpen
  \bibfield  {author} {\bibinfo {author} {\bibfnamefont {A.~S.}\ \bibnamefont {Alexandrov}}\ and\ \bibinfo {author} {\bibfnamefont {J.~T.}\ \bibnamefont {Devreese}},\ }\href {https://link.springer.com/book/10.1007/978-3-642-01896-1} {\emph {\bibinfo {title} {Advances in polaron physics}}},\ Vol.\ \bibinfo {volume} {159}\ (\bibinfo  {publisher} {Springer},\ \bibinfo {year} {2010})\BibitemShut {NoStop}%
\bibitem [{\citenamefont {Landau}(1933)}]{landau1933electron}%
  \BibitemOpen
  \bibfield  {author} {\bibinfo {author} {\bibfnamefont {L.~D.}\ \bibnamefont {Landau}},\ }\bibfield  {title} {\bibinfo {title} {Electron motion in crystal lattices},\ }\href@noop {} {\bibfield  {journal} {\bibinfo  {journal} {Phys. Z. Sowjetunion}\ }\textbf {\bibinfo {volume} {3}},\ \bibinfo {pages} {664} (\bibinfo {year} {1933})}\BibitemShut {NoStop}%
\bibitem [{\citenamefont {Pekar}(1946)}]{pekar1946local}%
  \BibitemOpen
  \bibfield  {author} {\bibinfo {author} {\bibfnamefont {S.~I.}\ \bibnamefont {Pekar}},\ }\bibfield  {title} {\bibinfo {title} {Local quantum states of electrons in an ideal ion crystal},\ }\href@noop {} {\bibfield  {journal} {\bibinfo  {journal} {Zh. Eksp. Teor. Fiz}\ }\textbf {\bibinfo {volume} {16}},\ \bibinfo {pages} {341} (\bibinfo {year} {1946})}\BibitemShut {NoStop}%
\bibitem [{\citenamefont {Landau}\ and\ \citenamefont {Pekar}(1948)}]{landau1948effective}%
  \BibitemOpen
  \bibfield  {author} {\bibinfo {author} {\bibfnamefont {L.~D.}\ \bibnamefont {Landau}}\ and\ \bibinfo {author} {\bibfnamefont {S.~I.}\ \bibnamefont {Pekar}},\ }\bibfield  {title} {\bibinfo {title} {The effective mass of the polaron},\ }\href@noop {} {\bibfield  {journal} {\bibinfo  {journal} {Zh. Eksp. Teor. Fiz}\ }\textbf {\bibinfo {volume} {18}},\ \bibinfo {pages} {419} (\bibinfo {year} {1948})}\BibitemShut {NoStop}%
\bibitem [{\citenamefont {Fr{\"o}hlich}\ \emph {et~al.}(1950)\citenamefont {Fr{\"o}hlich}, \citenamefont {Pelzer},\ and\ \citenamefont {Zienau}}]{frohlich1950xx}%
  \BibitemOpen
  \bibfield  {author} {\bibinfo {author} {\bibfnamefont {H.}~\bibnamefont {Fr{\"o}hlich}}, \bibinfo {author} {\bibfnamefont {H.}~\bibnamefont {Pelzer}},\ and\ \bibinfo {author} {\bibfnamefont {S.}~\bibnamefont {Zienau}},\ }\bibfield  {title} {\bibinfo {title} {Xx. properties of slow electrons in polar materials},\ }\href {https://www.tandfonline.com/doi/abs/10.1080/14786445008521794} {\bibfield  {journal} {\bibinfo  {journal} {The London, Edinburgh, and Dublin Philosophical Magazine and Journal of Science}\ }\textbf {\bibinfo {volume} {41}},\ \bibinfo {pages} {221} (\bibinfo {year} {1950})}\BibitemShut {NoStop}%
\bibitem [{\citenamefont {Lee}\ \emph {et~al.}(1953)\citenamefont {Lee}, \citenamefont {Low},\ and\ \citenamefont {Pines}}]{lee1953motion}%
  \BibitemOpen
  \bibfield  {author} {\bibinfo {author} {\bibfnamefont {T.}~\bibnamefont {Lee}}, \bibinfo {author} {\bibfnamefont {F.}~\bibnamefont {Low}},\ and\ \bibinfo {author} {\bibfnamefont {D.}~\bibnamefont {Pines}},\ }\bibfield  {title} {\bibinfo {title} {The motion of slow electrons in a polar crystal},\ }\href {https://journals.aps.org/pr/abstract/10.1103/PhysRev.90.297} {\bibfield  {journal} {\bibinfo  {journal} {Physical Review}\ }\textbf {\bibinfo {volume} {90}},\ \bibinfo {pages} {297} (\bibinfo {year} {1953})}\BibitemShut {NoStop}%
\bibitem [{\citenamefont {Feynman}(1955)}]{feynman1955slow}%
  \BibitemOpen
  \bibfield  {author} {\bibinfo {author} {\bibfnamefont {R.~P.}\ \bibnamefont {Feynman}},\ }\bibfield  {title} {\bibinfo {title} {Slow electrons in a polar crystal},\ }\href {https://journals.aps.org/pr/abstract/10.1103/PhysRev.97.660} {\bibfield  {journal} {\bibinfo  {journal} {Physical Review}\ }\textbf {\bibinfo {volume} {97}},\ \bibinfo {pages} {660} (\bibinfo {year} {1955})}\BibitemShut {NoStop}%
\bibitem [{\citenamefont {Devreese}\ and\ \citenamefont {Alexandrov}(2009)}]{devreese2009frohlich}%
  \BibitemOpen
  \bibfield  {author} {\bibinfo {author} {\bibfnamefont {J.~T.}\ \bibnamefont {Devreese}}\ and\ \bibinfo {author} {\bibfnamefont {A.~S.}\ \bibnamefont {Alexandrov}},\ }\bibfield  {title} {\bibinfo {title} {Fr{\"o}hlich polaron and bipolaron: recent developments},\ }\href {https://iopscience.iop.org/article/10.1088/0034-4885/72/6/066501} {\bibfield  {journal} {\bibinfo  {journal} {Reports on Progress in Physics}\ }\textbf {\bibinfo {volume} {72}},\ \bibinfo {pages} {066501} (\bibinfo {year} {2009})}\BibitemShut {NoStop}%
\bibitem [{\citenamefont {Franchini}\ \emph {et~al.}(2021)\citenamefont {Franchini}, \citenamefont {Reticcioli}, \citenamefont {Setvin},\ and\ \citenamefont {Diebold}}]{franchini2021polarons}%
  \BibitemOpen
  \bibfield  {author} {\bibinfo {author} {\bibfnamefont {C.}~\bibnamefont {Franchini}}, \bibinfo {author} {\bibfnamefont {M.}~\bibnamefont {Reticcioli}}, \bibinfo {author} {\bibfnamefont {M.}~\bibnamefont {Setvin}},\ and\ \bibinfo {author} {\bibfnamefont {U.}~\bibnamefont {Diebold}},\ }\bibfield  {title} {\bibinfo {title} {Polarons in materials},\ }\href@noop {} {\bibfield  {journal} {\bibinfo  {journal} {Nature Reviews Materials}\ }\textbf {\bibinfo {volume} {6}},\ \bibinfo {pages} {560} (\bibinfo {year} {2021})}\BibitemShut {NoStop}%
\bibitem [{\citenamefont {Prokof'ev}\ and\ \citenamefont {Svistunov}(1998)}]{prokofev1998polaron}%
  \BibitemOpen
  \bibfield  {author} {\bibinfo {author} {\bibfnamefont {N.~V.}\ \bibnamefont {Prokof'ev}}\ and\ \bibinfo {author} {\bibfnamefont {B.~V.}\ \bibnamefont {Svistunov}},\ }\bibfield  {title} {\bibinfo {title} {Polaron problem by diagrammatic quantum monte carlo},\ }\href {https://journals.aps.org/prl/abstract/10.1103/PhysRevLett.81.2514} {\bibfield  {journal} {\bibinfo  {journal} {Physical review letters}\ }\textbf {\bibinfo {volume} {81}},\ \bibinfo {pages} {2514} (\bibinfo {year} {1998})}\BibitemShut {NoStop}%
\bibitem [{\citenamefont {Mishchenko}\ \emph {et~al.}(2000)\citenamefont {Mishchenko}, \citenamefont {Prokof’Ev}, \citenamefont {Sakamoto},\ and\ \citenamefont {Svistunov}}]{mishchenko2000diagrammatic}%
  \BibitemOpen
  \bibfield  {author} {\bibinfo {author} {\bibfnamefont {A.~S.}\ \bibnamefont {Mishchenko}}, \bibinfo {author} {\bibfnamefont {N.~V.}\ \bibnamefont {Prokof’Ev}}, \bibinfo {author} {\bibfnamefont {A.}~\bibnamefont {Sakamoto}},\ and\ \bibinfo {author} {\bibfnamefont {B.~V.}\ \bibnamefont {Svistunov}},\ }\bibfield  {title} {\bibinfo {title} {Diagrammatic quantum monte carlo study of the fr{\"o}hlich polaron},\ }\href {https://journals.aps.org/prb/abstract/10.1103/PhysRevB.62.6317} {\bibfield  {journal} {\bibinfo  {journal} {Physical Review B}\ }\textbf {\bibinfo {volume} {62}},\ \bibinfo {pages} {6317} (\bibinfo {year} {2000})}\BibitemShut {NoStop}%
\bibitem [{\citenamefont {Hahn}\ \emph {et~al.}(2018)\citenamefont {Hahn}, \citenamefont {Klimin}, \citenamefont {Tempere}, \citenamefont {Devreese},\ and\ \citenamefont {Franchini}}]{hahn2018diagrammatic}%
  \BibitemOpen
  \bibfield  {author} {\bibinfo {author} {\bibfnamefont {T.}~\bibnamefont {Hahn}}, \bibinfo {author} {\bibfnamefont {S.}~\bibnamefont {Klimin}}, \bibinfo {author} {\bibfnamefont {J.}~\bibnamefont {Tempere}}, \bibinfo {author} {\bibfnamefont {J.~T.}\ \bibnamefont {Devreese}},\ and\ \bibinfo {author} {\bibfnamefont {C.}~\bibnamefont {Franchini}},\ }\bibfield  {title} {\bibinfo {title} {Diagrammatic monte carlo study of fr{\"o}hlich polaron dispersion in two and three dimensions},\ }\href {https://journals.aps.org/prb/abstract/10.1103/PhysRevB.97.134305} {\bibfield  {journal} {\bibinfo  {journal} {Physical Review B}\ }\textbf {\bibinfo {volume} {97}},\ \bibinfo {pages} {134305} (\bibinfo {year} {2018})}\BibitemShut {NoStop}%
\bibitem [{\citenamefont {Giustino}(2017)}]{giustino2017electron}%
  \BibitemOpen
  \bibfield  {author} {\bibinfo {author} {\bibfnamefont {F.}~\bibnamefont {Giustino}},\ }\bibfield  {title} {\bibinfo {title} {Electron-phonon interactions from first principles},\ }\href {https://journals.aps.org/rmp/abstract/10.1103/RevModPhys.89.015003} {\bibfield  {journal} {\bibinfo  {journal} {Reviews of Modern Physics}\ }\textbf {\bibinfo {volume} {89}},\ \bibinfo {pages} {015003} (\bibinfo {year} {2017})}\BibitemShut {NoStop}%
\bibitem [{\citenamefont {Sio}\ \emph {et~al.}(2019{\natexlab{a}})\citenamefont {Sio}, \citenamefont {Verdi}, \citenamefont {Ponc{\'e}},\ and\ \citenamefont {Giustino}}]{sio2019polarons}%
  \BibitemOpen
  \bibfield  {author} {\bibinfo {author} {\bibfnamefont {W.~H.}\ \bibnamefont {Sio}}, \bibinfo {author} {\bibfnamefont {C.}~\bibnamefont {Verdi}}, \bibinfo {author} {\bibfnamefont {S.}~\bibnamefont {Ponc{\'e}}},\ and\ \bibinfo {author} {\bibfnamefont {F.}~\bibnamefont {Giustino}},\ }\bibfield  {title} {\bibinfo {title} {Polarons from first principles, without supercells},\ }\href {https://journals.aps.org/prl/abstract/10.1103/PhysRevLett.122.246403} {\bibfield  {journal} {\bibinfo  {journal} {Physical Review Letters}\ }\textbf {\bibinfo {volume} {122}},\ \bibinfo {pages} {246403} (\bibinfo {year} {2019}{\natexlab{a}})}\BibitemShut {NoStop}%
\bibitem [{\citenamefont {Sio}\ \emph {et~al.}(2019{\natexlab{b}})\citenamefont {Sio}, \citenamefont {Verdi}, \citenamefont {Ponc{\'e}},\ and\ \citenamefont {Giustino}}]{sio2019ab}%
  \BibitemOpen
  \bibfield  {author} {\bibinfo {author} {\bibfnamefont {W.~H.}\ \bibnamefont {Sio}}, \bibinfo {author} {\bibfnamefont {C.}~\bibnamefont {Verdi}}, \bibinfo {author} {\bibfnamefont {S.}~\bibnamefont {Ponc{\'e}}},\ and\ \bibinfo {author} {\bibfnamefont {F.}~\bibnamefont {Giustino}},\ }\bibfield  {title} {\bibinfo {title} {Ab initio theory of polarons: Formalism and applications},\ }\href {https://journals.aps.org/prb/abstract/10.1103/PhysRevB.99.235139} {\bibfield  {journal} {\bibinfo  {journal} {Physical Review B}\ }\textbf {\bibinfo {volume} {99}},\ \bibinfo {pages} {235139} (\bibinfo {year} {2019}{\natexlab{b}})}\BibitemShut {NoStop}%
\bibitem [{\citenamefont {Haken}(1958)}]{haken1958theorie}%
  \BibitemOpen
  \bibfield  {author} {\bibinfo {author} {\bibfnamefont {H.}~\bibnamefont {Haken}},\ }\bibfield  {title} {\bibinfo {title} {Die theorie des exzitons im festen k{\"o}rper},\ }\href {https://onlinelibrary.wiley.com/doi/10.1002/prop.19580060602} {\bibfield  {journal} {\bibinfo  {journal} {Fortschritte der Physik}\ }\textbf {\bibinfo {volume} {6}},\ \bibinfo {pages} {271} (\bibinfo {year} {1958})}\BibitemShut {NoStop}%
\bibitem [{\citenamefont {Mahanti}\ and\ \citenamefont {Varma}(1972)}]{mahanti1972effective}%
  \BibitemOpen
  \bibfield  {author} {\bibinfo {author} {\bibfnamefont {S.}~\bibnamefont {Mahanti}}\ and\ \bibinfo {author} {\bibfnamefont {C.}~\bibnamefont {Varma}},\ }\bibfield  {title} {\bibinfo {title} {Effective electron-hole interactions in polar semiconductors},\ }\href {https://journals.aps.org/prb/abstract/10.1103/PhysRevB.6.2209} {\bibfield  {journal} {\bibinfo  {journal} {Physical Review B}\ }\textbf {\bibinfo {volume} {6}},\ \bibinfo {pages} {2209} (\bibinfo {year} {1972})}\BibitemShut {NoStop}%
\bibitem [{\citenamefont {Bajaj}(1974)}]{bajaj1974effect}%
  \BibitemOpen
  \bibfield  {author} {\bibinfo {author} {\bibfnamefont {K.}~\bibnamefont {Bajaj}},\ }\bibfield  {title} {\bibinfo {title} {Effect of electron-phonon interaction on the binding energy of a wannier exciton in a polarizable medium},\ }\href {https://www.sciencedirect.com/science/article/abs/pii/0038109874900556} {\bibfield  {journal} {\bibinfo  {journal} {Solid State Communications}\ }\textbf {\bibinfo {volume} {15}},\ \bibinfo {pages} {1221} (\bibinfo {year} {1974})}\BibitemShut {NoStop}%
\bibitem [{\citenamefont {Pollmann}\ and\ \citenamefont {B{\"u}ttner}(1977)}]{pollmann1977effective}%
  \BibitemOpen
  \bibfield  {author} {\bibinfo {author} {\bibfnamefont {J.}~\bibnamefont {Pollmann}}\ and\ \bibinfo {author} {\bibfnamefont {H.}~\bibnamefont {B{\"u}ttner}},\ }\bibfield  {title} {\bibinfo {title} {Effective hamiltonians and bindings energies of wannier excitons in polar semiconductors},\ }\href {https://journals.aps.org/prb/abstract/10.1103/PhysRevB.16.4480} {\bibfield  {journal} {\bibinfo  {journal} {Physical Review B}\ }\textbf {\bibinfo {volume} {16}},\ \bibinfo {pages} {4480} (\bibinfo {year} {1977})}\BibitemShut {NoStop}%
\bibitem [{\citenamefont {Kane}(1978)}]{kane1978pollmann}%
  \BibitemOpen
  \bibfield  {author} {\bibinfo {author} {\bibfnamefont {E.~O.}\ \bibnamefont {Kane}},\ }\bibfield  {title} {\bibinfo {title} {Pollmann-b{\"u}ttner variational method for excitonic polarons},\ }\href {https://journals.aps.org/prb/abstract/10.1103/PhysRevB.18.6849} {\bibfield  {journal} {\bibinfo  {journal} {Physical Review B}\ }\textbf {\bibinfo {volume} {18}},\ \bibinfo {pages} {6849} (\bibinfo {year} {1978})}\BibitemShut {NoStop}%
\bibitem [{\citenamefont {Burovski}\ \emph {et~al.}(2001)\citenamefont {Burovski}, \citenamefont {Mishchenko}, \citenamefont {Prokof'ev},\ and\ \citenamefont {Svistunov}}]{burovski2001diagrammatic}%
  \BibitemOpen
  \bibfield  {author} {\bibinfo {author} {\bibfnamefont {E.~A.}\ \bibnamefont {Burovski}}, \bibinfo {author} {\bibfnamefont {A.~S.}\ \bibnamefont {Mishchenko}}, \bibinfo {author} {\bibfnamefont {N.~V.}\ \bibnamefont {Prokof'ev}},\ and\ \bibinfo {author} {\bibfnamefont {B.~V.}\ \bibnamefont {Svistunov}},\ }\bibfield  {title} {\bibinfo {title} {Diagrammatic quantum monte carlo for two-body problems: Applied to excitons},\ }\href {https://journals.aps.org/prl/abstract/10.1103/PhysRevLett.87.186402} {\bibfield  {journal} {\bibinfo  {journal} {Physical Review Letters}\ }\textbf {\bibinfo {volume} {87}},\ \bibinfo {pages} {186402} (\bibinfo {year} {2001})}\BibitemShut {NoStop}%
\bibitem [{\citenamefont {Burovski}\ \emph {et~al.}(2008)\citenamefont {Burovski}, \citenamefont {Fehske},\ and\ \citenamefont {Mishchenko}}]{burovski2008exact}%
  \BibitemOpen
  \bibfield  {author} {\bibinfo {author} {\bibfnamefont {E.}~\bibnamefont {Burovski}}, \bibinfo {author} {\bibfnamefont {H.}~\bibnamefont {Fehske}},\ and\ \bibinfo {author} {\bibfnamefont {A.~S.}\ \bibnamefont {Mishchenko}},\ }\bibfield  {title} {\bibinfo {title} {Exact treatment of exciton-polaron formation by diagrammatic monte carlo simulations},\ }\href {https://journals.aps.org/prl/abstract/10.1103/PhysRevLett.101.116403} {\bibfield  {journal} {\bibinfo  {journal} {Physical Review Letters}\ }\textbf {\bibinfo {volume} {101}},\ \bibinfo {pages} {116403} (\bibinfo {year} {2008})}\BibitemShut {NoStop}%
\bibitem [{\citenamefont {Mishchenko}\ \emph {et~al.}(2018)\citenamefont {Mishchenko}, \citenamefont {De~Filippis}, \citenamefont {Cataudella}, \citenamefont {Nagaosa},\ and\ \citenamefont {Fehske}}]{mishchenko2018optical}%
  \BibitemOpen
  \bibfield  {author} {\bibinfo {author} {\bibfnamefont {A.~S.}\ \bibnamefont {Mishchenko}}, \bibinfo {author} {\bibfnamefont {G.}~\bibnamefont {De~Filippis}}, \bibinfo {author} {\bibfnamefont {V.}~\bibnamefont {Cataudella}}, \bibinfo {author} {\bibfnamefont {N.}~\bibnamefont {Nagaosa}},\ and\ \bibinfo {author} {\bibfnamefont {H.}~\bibnamefont {Fehske}},\ }\bibfield  {title} {\bibinfo {title} {Optical signatures of exciton polarons from diagrammatic monte carlo},\ }\href {https://journals.aps.org/prb/abstract/10.1103/PhysRevB.97.045141} {\bibfield  {journal} {\bibinfo  {journal} {Physical Review B}\ }\textbf {\bibinfo {volume} {97}},\ \bibinfo {pages} {045141} (\bibinfo {year} {2018})}\BibitemShut {NoStop}%
\bibitem [{\citenamefont {Rana}\ and\ \citenamefont {Limmer}(2025)}]{rana2024interplay}%
  \BibitemOpen
  \bibfield  {author} {\bibinfo {author} {\bibfnamefont {R.}~\bibnamefont {Rana}}\ and\ \bibinfo {author} {\bibfnamefont {D.~T.}\ \bibnamefont {Limmer}},\ }\bibfield  {title} {\bibinfo {title} {On the interplay of electronic and lattice screening on exciton binding in two-dimensional lead halide perovskites},\ }\href {https://pubs.acs.org/doi/10.1021/acs.nanolett.4c05646} {\bibfield  {journal} {\bibinfo  {journal} {Nano Letters}\ }\textbf {\bibinfo {volume} {25}},\ \bibinfo {pages} {4727–4733} (\bibinfo {year} {2025})}\BibitemShut {NoStop}%
\bibitem [{\citenamefont {Filip}\ \emph {et~al.}(2021)\citenamefont {Filip}, \citenamefont {Haber},\ and\ \citenamefont {Neaton}}]{filip2021phonon}%
  \BibitemOpen
  \bibfield  {author} {\bibinfo {author} {\bibfnamefont {M.~R.}\ \bibnamefont {Filip}}, \bibinfo {author} {\bibfnamefont {J.~B.}\ \bibnamefont {Haber}},\ and\ \bibinfo {author} {\bibfnamefont {J.~B.}\ \bibnamefont {Neaton}},\ }\bibfield  {title} {\bibinfo {title} {Phonon screening of excitons in semiconductors: halide perovskites and beyond},\ }\href {https://journals.aps.org/prl/abstract/10.1103/PhysRevLett.127.067401} {\bibfield  {journal} {\bibinfo  {journal} {Physical review letters}\ }\textbf {\bibinfo {volume} {127}},\ \bibinfo {pages} {067401} (\bibinfo {year} {2021})}\BibitemShut {NoStop}%
\bibitem [{\citenamefont {Alvertis}\ \emph {et~al.}(2024)\citenamefont {Alvertis}, \citenamefont {Haber}, \citenamefont {Li}, \citenamefont {Coveney}, \citenamefont {Louie}, \citenamefont {Filip},\ and\ \citenamefont {Neaton}}]{alvertis2024phonon}%
  \BibitemOpen
  \bibfield  {author} {\bibinfo {author} {\bibfnamefont {A.~M.}\ \bibnamefont {Alvertis}}, \bibinfo {author} {\bibfnamefont {J.~B.}\ \bibnamefont {Haber}}, \bibinfo {author} {\bibfnamefont {Z.}~\bibnamefont {Li}}, \bibinfo {author} {\bibfnamefont {C.~J.}\ \bibnamefont {Coveney}}, \bibinfo {author} {\bibfnamefont {S.~G.}\ \bibnamefont {Louie}}, \bibinfo {author} {\bibfnamefont {M.~R.}\ \bibnamefont {Filip}},\ and\ \bibinfo {author} {\bibfnamefont {J.~B.}\ \bibnamefont {Neaton}},\ }\bibfield  {title} {\bibinfo {title} {Phonon screening and dissociation of excitons at finite temperatures from first principles},\ }\href {https://www.pnas.org/doi/10.1073/pnas.2403434121} {\bibfield  {journal} {\bibinfo  {journal} {Proceedings of the National Academy of Sciences}\ }\textbf {\bibinfo {volume} {121}},\ \bibinfo {pages} {e2403434121} (\bibinfo {year} {2024})}\BibitemShut {NoStop}%
\bibitem [{\citenamefont {Dai}\ \emph {et~al.}(2024)\citenamefont {Dai}, \citenamefont {Lian}, \citenamefont {Lafuente-Bartolome},\ and\ \citenamefont {Giustino}}]{dai2024excitonic}%
  \BibitemOpen
  \bibfield  {author} {\bibinfo {author} {\bibfnamefont {Z.}~\bibnamefont {Dai}}, \bibinfo {author} {\bibfnamefont {C.}~\bibnamefont {Lian}}, \bibinfo {author} {\bibfnamefont {J.}~\bibnamefont {Lafuente-Bartolome}},\ and\ \bibinfo {author} {\bibfnamefont {F.}~\bibnamefont {Giustino}},\ }\bibfield  {title} {\bibinfo {title} {Excitonic polarons and self-trapped excitons from first-principles exciton-phonon couplings},\ }\href {https://journals.aps.org/prl/abstract/10.1103/PhysRevLett.132.036902} {\bibfield  {journal} {\bibinfo  {journal} {Physical Review Letters}\ }\textbf {\bibinfo {volume} {132}},\ \bibinfo {pages} {036902} (\bibinfo {year} {2024})}\BibitemShut {NoStop}%
\bibitem [{\citenamefont {Baranowski}\ and\ \citenamefont {Plochocka}(2020)}]{baranowski2020excitons}%
  \BibitemOpen
  \bibfield  {author} {\bibinfo {author} {\bibfnamefont {M.}~\bibnamefont {Baranowski}}\ and\ \bibinfo {author} {\bibfnamefont {P.}~\bibnamefont {Plochocka}},\ }\bibfield  {title} {\bibinfo {title} {Excitons in metal-halide perovskites},\ }\href {https://onlinelibrary.wiley.com/doi/full/10.1002/aenm.201903659} {\bibfield  {journal} {\bibinfo  {journal} {Adv. Energy Mater.}\ }\textbf {\bibinfo {volume} {10}},\ \bibinfo {pages} {1903659} (\bibinfo {year} {2020})}\BibitemShut {NoStop}%
\bibitem [{\citenamefont {Simbula}\ \emph {et~al.}(2021)\citenamefont {Simbula}, \citenamefont {Pau}, \citenamefont {Wang}, \citenamefont {Liu}, \citenamefont {Sarritzu}, \citenamefont {Lai}, \citenamefont {Lodde}, \citenamefont {Mattana}, \citenamefont {Mula}, \citenamefont {Geddo~Lehmann} \emph {et~al.}}]{simbula2021polaron}%
  \BibitemOpen
  \bibfield  {author} {\bibinfo {author} {\bibfnamefont {A.}~\bibnamefont {Simbula}}, \bibinfo {author} {\bibfnamefont {R.}~\bibnamefont {Pau}}, \bibinfo {author} {\bibfnamefont {Q.}~\bibnamefont {Wang}}, \bibinfo {author} {\bibfnamefont {F.}~\bibnamefont {Liu}}, \bibinfo {author} {\bibfnamefont {V.}~\bibnamefont {Sarritzu}}, \bibinfo {author} {\bibfnamefont {S.}~\bibnamefont {Lai}}, \bibinfo {author} {\bibfnamefont {M.}~\bibnamefont {Lodde}}, \bibinfo {author} {\bibfnamefont {F.}~\bibnamefont {Mattana}}, \bibinfo {author} {\bibfnamefont {G.}~\bibnamefont {Mula}}, \bibinfo {author} {\bibfnamefont {A.}~\bibnamefont {Geddo~Lehmann}}, \emph {et~al.},\ }\bibfield  {title} {\bibinfo {title} {Polaron plasma in equilibrium with bright excitons in 2d and 3d hybrid perovskites},\ }\href {https://onlinelibrary.wiley.com/doi/full/10.1002/adom.202100295} {\bibfield  {journal} {\bibinfo  {journal} {Advanced Optical Materials}\ }\textbf {\bibinfo {volume} {9}},\ \bibinfo {pages} {2100295} (\bibinfo {year}
  {2021})}\BibitemShut {NoStop}%
\bibitem [{\citenamefont {Tao}\ \emph {et~al.}(2021)\citenamefont {Tao}, \citenamefont {Zhang}, \citenamefont {Zhou}, \citenamefont {Zhao},\ and\ \citenamefont {Zhu}}]{tao2021momentarily}%
  \BibitemOpen
  \bibfield  {author} {\bibinfo {author} {\bibfnamefont {W.}~\bibnamefont {Tao}}, \bibinfo {author} {\bibfnamefont {C.}~\bibnamefont {Zhang}}, \bibinfo {author} {\bibfnamefont {Q.}~\bibnamefont {Zhou}}, \bibinfo {author} {\bibfnamefont {Y.}~\bibnamefont {Zhao}},\ and\ \bibinfo {author} {\bibfnamefont {H.}~\bibnamefont {Zhu}},\ }\bibfield  {title} {\bibinfo {title} {Momentarily trapped exciton polaron in two-dimensional lead halide perovskites},\ }\href {https://www.nature.com/articles/s41467-021-21721-3} {\bibfield  {journal} {\bibinfo  {journal} {Nature communications}\ }\textbf {\bibinfo {volume} {12}},\ \bibinfo {pages} {1400} (\bibinfo {year} {2021})}\BibitemShut {NoStop}%
\bibitem [{\citenamefont {Tao}\ \emph {et~al.}(2022)\citenamefont {Tao}, \citenamefont {Zhang},\ and\ \citenamefont {Zhu}}]{tao2022dynamic}%
  \BibitemOpen
  \bibfield  {author} {\bibinfo {author} {\bibfnamefont {W.}~\bibnamefont {Tao}}, \bibinfo {author} {\bibfnamefont {Y.}~\bibnamefont {Zhang}},\ and\ \bibinfo {author} {\bibfnamefont {H.}~\bibnamefont {Zhu}},\ }\bibfield  {title} {\bibinfo {title} {Dynamic exciton polaron in two-dimensional lead halide perovskites and implications for optoelectronic applications},\ }\href {https://pubs.acs.org/doi/abs/10.1021/acs.accounts.1c00626} {\bibfield  {journal} {\bibinfo  {journal} {Accounts of chemical research}\ }\textbf {\bibinfo {volume} {55}},\ \bibinfo {pages} {345} (\bibinfo {year} {2022})}\BibitemShut {NoStop}%
\bibitem [{\citenamefont {Dyksik}\ \emph {et~al.}(2024)\citenamefont {Dyksik}, \citenamefont {Beret}, \citenamefont {Baranowski}, \citenamefont {Duim}, \citenamefont {Moyano}, \citenamefont {Posmyk}, \citenamefont {Mlayah}, \citenamefont {Adjokatse}, \citenamefont {Maude}, \citenamefont {Loi} \emph {et~al.}}]{dyksik2024polaron}%
  \BibitemOpen
  \bibfield  {author} {\bibinfo {author} {\bibfnamefont {M.}~\bibnamefont {Dyksik}}, \bibinfo {author} {\bibfnamefont {D.}~\bibnamefont {Beret}}, \bibinfo {author} {\bibfnamefont {M.}~\bibnamefont {Baranowski}}, \bibinfo {author} {\bibfnamefont {H.}~\bibnamefont {Duim}}, \bibinfo {author} {\bibfnamefont {S.}~\bibnamefont {Moyano}}, \bibinfo {author} {\bibfnamefont {K.}~\bibnamefont {Posmyk}}, \bibinfo {author} {\bibfnamefont {A.}~\bibnamefont {Mlayah}}, \bibinfo {author} {\bibfnamefont {S.}~\bibnamefont {Adjokatse}}, \bibinfo {author} {\bibfnamefont {D.~K.}\ \bibnamefont {Maude}}, \bibinfo {author} {\bibfnamefont {M.~A.}\ \bibnamefont {Loi}}, \emph {et~al.},\ }\bibfield  {title} {\bibinfo {title} {Polaron vibronic progression shapes the optical response of 2d perovskites},\ }\href {https://advanced.onlinelibrary.wiley.com/doi/full/10.1002/advs.202305182} {\bibfield  {journal} {\bibinfo  {journal} {Advanced Science}\ }\textbf {\bibinfo {volume} {11}},\ \bibinfo {pages} {2305182} (\bibinfo {year}
  {2024})}\BibitemShut {NoStop}%
\bibitem [{\citenamefont {Biswas}\ \emph {et~al.}(2024)\citenamefont {Biswas}, \citenamefont {Zhao}, \citenamefont {Alowa}, \citenamefont {Zacharias}, \citenamefont {Sharifzadeh}, \citenamefont {Coker}, \citenamefont {Seferos},\ and\ \citenamefont {Scholes}}]{biswas2024exciton}%
  \BibitemOpen
  \bibfield  {author} {\bibinfo {author} {\bibfnamefont {S.}~\bibnamefont {Biswas}}, \bibinfo {author} {\bibfnamefont {R.}~\bibnamefont {Zhao}}, \bibinfo {author} {\bibfnamefont {F.}~\bibnamefont {Alowa}}, \bibinfo {author} {\bibfnamefont {M.}~\bibnamefont {Zacharias}}, \bibinfo {author} {\bibfnamefont {S.}~\bibnamefont {Sharifzadeh}}, \bibinfo {author} {\bibfnamefont {D.~F.}\ \bibnamefont {Coker}}, \bibinfo {author} {\bibfnamefont {D.~S.}\ \bibnamefont {Seferos}},\ and\ \bibinfo {author} {\bibfnamefont {G.~D.}\ \bibnamefont {Scholes}},\ }\bibfield  {title} {\bibinfo {title} {Exciton polaron formation and hot-carrier relaxation in rigid dion--jacobson-type two-dimensional perovskites},\ }\href {https://www.nature.com/articles/s41563-024-01895-z} {\bibfield  {journal} {\bibinfo  {journal} {Nature Materials}\ ,\ \bibinfo {pages} {1}} (\bibinfo {year} {2024})}\BibitemShut {NoStop}%
\bibitem [{\citenamefont {Baranowski}\ \emph {et~al.}(2024)\citenamefont {Baranowski}, \citenamefont {Nowok}, \citenamefont {Galkowski}, \citenamefont {Dyksik}, \citenamefont {Surrente}, \citenamefont {Maude}, \citenamefont {Zacharias}, \citenamefont {Volonakis}, \citenamefont {Stranks}, \citenamefont {Even} \emph {et~al.}}]{baranowski2024polaronic}%
  \BibitemOpen
  \bibfield  {author} {\bibinfo {author} {\bibfnamefont {M.}~\bibnamefont {Baranowski}}, \bibinfo {author} {\bibfnamefont {A.}~\bibnamefont {Nowok}}, \bibinfo {author} {\bibfnamefont {K.}~\bibnamefont {Galkowski}}, \bibinfo {author} {\bibfnamefont {M.}~\bibnamefont {Dyksik}}, \bibinfo {author} {\bibfnamefont {A.}~\bibnamefont {Surrente}}, \bibinfo {author} {\bibfnamefont {D.}~\bibnamefont {Maude}}, \bibinfo {author} {\bibfnamefont {M.}~\bibnamefont {Zacharias}}, \bibinfo {author} {\bibfnamefont {G.}~\bibnamefont {Volonakis}}, \bibinfo {author} {\bibfnamefont {S.~D.}\ \bibnamefont {Stranks}}, \bibinfo {author} {\bibfnamefont {J.}~\bibnamefont {Even}}, \emph {et~al.},\ }\bibfield  {title} {\bibinfo {title} {Polaronic mass enhancement and polaronic excitons in metal halide perovskites},\ }\href {https://pubs.acs.org/doi/full/10.1021/acsenergylett.4c00905} {\bibfield  {journal} {\bibinfo  {journal} {ACS Energy Letters}\ }\textbf {\bibinfo {volume} {9}},\ \bibinfo {pages} {2696} (\bibinfo {year}
  {2024})}\BibitemShut {NoStop}%
\bibitem [{\citenamefont {Lei}\ \emph {et~al.}(2024)\citenamefont {Lei}, \citenamefont {Xu}, \citenamefont {Zhang}, \citenamefont {Feng}, \citenamefont {Zhou}, \citenamefont {Tang}, \citenamefont {Wang}, \citenamefont {Li}, \citenamefont {Nan}, \citenamefont {Xu} \emph {et~al.}}]{lei2024persistent}%
  \BibitemOpen
  \bibfield  {author} {\bibinfo {author} {\bibfnamefont {H.}~\bibnamefont {Lei}}, \bibinfo {author} {\bibfnamefont {Y.}~\bibnamefont {Xu}}, \bibinfo {author} {\bibfnamefont {Y.}~\bibnamefont {Zhang}}, \bibinfo {author} {\bibfnamefont {Q.}~\bibnamefont {Feng}}, \bibinfo {author} {\bibfnamefont {H.}~\bibnamefont {Zhou}}, \bibinfo {author} {\bibfnamefont {W.}~\bibnamefont {Tang}}, \bibinfo {author} {\bibfnamefont {J.}~\bibnamefont {Wang}}, \bibinfo {author} {\bibfnamefont {L.}~\bibnamefont {Li}}, \bibinfo {author} {\bibfnamefont {G.}~\bibnamefont {Nan}}, \bibinfo {author} {\bibfnamefont {W.}~\bibnamefont {Xu}}, \emph {et~al.},\ }\bibfield  {title} {\bibinfo {title} {Persistent exciton dressed by weak polaronic effect in rigid and harmonic lattice dion--jacobson 2d perovskites},\ }\href {https://pubs.acs.org/doi/abs/10.1021/acsnano.4c12132} {\bibfield  {journal} {\bibinfo  {journal} {ACS nano}\ }\textbf {\bibinfo {volume} {18}},\ \bibinfo {pages} {31485} (\bibinfo {year} {2024})}\BibitemShut {NoStop}%
\bibitem [{\citenamefont {Duan}\ \emph {et~al.}(2024)\citenamefont {Duan}, \citenamefont {Li}, \citenamefont {Divitini}, \citenamefont {Cortecchia}, \citenamefont {Yuan}, \citenamefont {You}, \citenamefont {Liu}, \citenamefont {Petrozza}, \citenamefont {Wu},\ and\ \citenamefont {Xi}}]{duan20242d}%
  \BibitemOpen
  \bibfield  {author} {\bibinfo {author} {\bibfnamefont {J.}~\bibnamefont {Duan}}, \bibinfo {author} {\bibfnamefont {J.}~\bibnamefont {Li}}, \bibinfo {author} {\bibfnamefont {G.}~\bibnamefont {Divitini}}, \bibinfo {author} {\bibfnamefont {D.}~\bibnamefont {Cortecchia}}, \bibinfo {author} {\bibfnamefont {F.}~\bibnamefont {Yuan}}, \bibinfo {author} {\bibfnamefont {J.}~\bibnamefont {You}}, \bibinfo {author} {\bibfnamefont {S.}~\bibnamefont {Liu}}, \bibinfo {author} {\bibfnamefont {A.}~\bibnamefont {Petrozza}}, \bibinfo {author} {\bibfnamefont {Z.}~\bibnamefont {Wu}},\ and\ \bibinfo {author} {\bibfnamefont {J.}~\bibnamefont {Xi}},\ }\bibfield  {title} {\bibinfo {title} {2d hybrid perovskites: From static and dynamic structures to potential applications},\ }\href {https://advanced.onlinelibrary.wiley.com/doi/full/10.1002/adma.202403455} {\bibfield  {journal} {\bibinfo  {journal} {Advanced Materials}\ }\textbf {\bibinfo {volume} {36}},\ \bibinfo {pages} {2403455} (\bibinfo {year} {2024})}\BibitemShut {NoStop}%
\bibitem [{\citenamefont {Novoselov}\ \emph {et~al.}(2016)\citenamefont {Novoselov}, \citenamefont {Mishchenko}, \citenamefont {Carvalho},\ and\ \citenamefont {Castro~Neto}}]{novoselov20162d}%
  \BibitemOpen
  \bibfield  {author} {\bibinfo {author} {\bibfnamefont {K.~S.}\ \bibnamefont {Novoselov}}, \bibinfo {author} {\bibfnamefont {A.}~\bibnamefont {Mishchenko}}, \bibinfo {author} {\bibfnamefont {A.}~\bibnamefont {Carvalho}},\ and\ \bibinfo {author} {\bibfnamefont {A.}~\bibnamefont {Castro~Neto}},\ }\bibfield  {title} {\bibinfo {title} {2d materials and van der waals heterostructures},\ }\href {https://www.science.org/doi/10.1126/science.aac9439} {\bibfield  {journal} {\bibinfo  {journal} {Science}\ }\textbf {\bibinfo {volume} {353}},\ \bibinfo {pages} {aac9439} (\bibinfo {year} {2016})}\BibitemShut {NoStop}%
\bibitem [{\citenamefont {Keldysh}(1979)}]{Keldysh1979}%
  \BibitemOpen
  \bibfield  {author} {\bibinfo {author} {\bibfnamefont {L.~V.}\ \bibnamefont {Keldysh}},\ }\bibfield  {title} {\bibinfo {title} {Coulomb interaction in thin semiconductor and semimetal films},\ }\href {http://jetpletters.ru/ps/0/article_22207.shtml} {\bibfield  {journal} {\bibinfo  {journal} {JETP Lett.}\ }\textbf {\bibinfo {volume} {29}},\ \bibinfo {pages} {658} (\bibinfo {year} {1979})}\BibitemShut {NoStop}%
\bibitem [{\citenamefont {Rytova}(1967)}]{Rytova1967}%
  \BibitemOpen
  \bibfield  {author} {\bibinfo {author} {\bibfnamefont {N.~S.}\ \bibnamefont {Rytova}},\ }\bibfield  {title} {\bibinfo {title} {The screened potential of a point charge in a thin film},\ }\href {http://vmu.phys.msu.ru/en/abstract/1967/3/1967-3-030/} {\bibfield  {journal} {\bibinfo  {journal} {Moscow University Physics Bulletin}\ }\textbf {\bibinfo {volume} {22}},\ \bibinfo {pages} {18} (\bibinfo {year} {1967})}\BibitemShut {NoStop}%
\bibitem [{\citenamefont {Cudazzo}\ \emph {et~al.}(2011)\citenamefont {Cudazzo}, \citenamefont {Tokatly},\ and\ \citenamefont {Rubio}}]{cudazzo2011dielectric}%
  \BibitemOpen
  \bibfield  {author} {\bibinfo {author} {\bibfnamefont {P.}~\bibnamefont {Cudazzo}}, \bibinfo {author} {\bibfnamefont {I.~V.}\ \bibnamefont {Tokatly}},\ and\ \bibinfo {author} {\bibfnamefont {A.}~\bibnamefont {Rubio}},\ }\bibfield  {title} {\bibinfo {title} {Dielectric screening in two-dimensional insulators: Implications for excitonic and impurity states in graphane},\ }\href {https://journals.aps.org/prb/abstract/10.1103/PhysRevB.84.085406} {\bibfield  {journal} {\bibinfo  {journal} {Physical Review B—Condensed Matter and Materials Physics}\ }\textbf {\bibinfo {volume} {84}},\ \bibinfo {pages} {085406} (\bibinfo {year} {2011})}\BibitemShut {NoStop}%
\bibitem [{\citenamefont {Sohier}\ \emph {et~al.}(2016)\citenamefont {Sohier}, \citenamefont {Calandra},\ and\ \citenamefont {Mauri}}]{sohier2016two}%
  \BibitemOpen
  \bibfield  {author} {\bibinfo {author} {\bibfnamefont {T.}~\bibnamefont {Sohier}}, \bibinfo {author} {\bibfnamefont {M.}~\bibnamefont {Calandra}},\ and\ \bibinfo {author} {\bibfnamefont {F.}~\bibnamefont {Mauri}},\ }\bibfield  {title} {\bibinfo {title} {Two-dimensional fr{\"o}hlich interaction in transition-metal dichalcogenide monolayers: Theoretical modeling and first-principles calculations},\ }\href {https://journals.aps.org/prb/abstract/10.1103/PhysRevB.94.085415} {\bibfield  {journal} {\bibinfo  {journal} {Physical Review B}\ }\textbf {\bibinfo {volume} {94}},\ \bibinfo {pages} {085415} (\bibinfo {year} {2016})}\BibitemShut {NoStop}%
\bibitem [{\citenamefont {Sohier}\ \emph {et~al.}(2017)\citenamefont {Sohier}, \citenamefont {Gibertini}, \citenamefont {Calandra}, \citenamefont {Mauri},\ and\ \citenamefont {Marzari}}]{sohier2017breakdown}%
  \BibitemOpen
  \bibfield  {author} {\bibinfo {author} {\bibfnamefont {T.}~\bibnamefont {Sohier}}, \bibinfo {author} {\bibfnamefont {M.}~\bibnamefont {Gibertini}}, \bibinfo {author} {\bibfnamefont {M.}~\bibnamefont {Calandra}}, \bibinfo {author} {\bibfnamefont {F.}~\bibnamefont {Mauri}},\ and\ \bibinfo {author} {\bibfnamefont {N.}~\bibnamefont {Marzari}},\ }\bibfield  {title} {\bibinfo {title} {Breakdown of optical phonons’ splitting in two-dimensional materials},\ }\href {https://pubs.acs.org/doi/10.1021/acs.nanolett.7b01090} {\bibfield  {journal} {\bibinfo  {journal} {Nano letters}\ }\textbf {\bibinfo {volume} {17}},\ \bibinfo {pages} {3758} (\bibinfo {year} {2017})}\BibitemShut {NoStop}%
\bibitem [{\citenamefont {Shahnazaryan}\ \emph {et~al.}(2025)\citenamefont {Shahnazaryan}, \citenamefont {Kudlis},\ and\ \citenamefont {Tokatly}}]{shahnazaryan2025polarons}%
  \BibitemOpen
  \bibfield  {author} {\bibinfo {author} {\bibfnamefont {V.}~\bibnamefont {Shahnazaryan}}, \bibinfo {author} {\bibfnamefont {A.}~\bibnamefont {Kudlis}},\ and\ \bibinfo {author} {\bibfnamefont {I.}~\bibnamefont {Tokatly}},\ }\bibfield  {title} {\bibinfo {title} {Polarons and exciton-polarons in two-dimensional polar materials},\ }\href {https://journals.aps.org/prl/abstract/10.1103/84p5-s6lj} {\bibfield  {journal} {\bibinfo  {journal} {Physical Review Letters}\ }\textbf {\bibinfo {volume} {135}},\ \bibinfo {pages} {066202} (\bibinfo {year} {2025})}\BibitemShut {NoStop}%
\bibitem [{\citenamefont {Li}\ \emph {et~al.}(2024{\natexlab{a}})\citenamefont {Li}, \citenamefont {Wang}, \citenamefont {Wang}, \citenamefont {Tao}, \citenamefont {Zhong}, \citenamefont {Su}, \citenamefont {Xue}, \citenamefont {Miao}, \citenamefont {Wang}, \citenamefont {Peng}, \citenamefont {Guo},\ and\ \citenamefont {Zhu}}]{Li2024experiment}%
  \BibitemOpen
  \bibfield  {author} {\bibinfo {author} {\bibfnamefont {J.}~\bibnamefont {Li}}, \bibinfo {author} {\bibfnamefont {L.}~\bibnamefont {Wang}}, \bibinfo {author} {\bibfnamefont {Y.}~\bibnamefont {Wang}}, \bibinfo {author} {\bibfnamefont {Z.}~\bibnamefont {Tao}}, \bibinfo {author} {\bibfnamefont {W.}~\bibnamefont {Zhong}}, \bibinfo {author} {\bibfnamefont {Z.}~\bibnamefont {Su}}, \bibinfo {author} {\bibfnamefont {S.}~\bibnamefont {Xue}}, \bibinfo {author} {\bibfnamefont {G.}~\bibnamefont {Miao}}, \bibinfo {author} {\bibfnamefont {W.}~\bibnamefont {Wang}}, \bibinfo {author} {\bibfnamefont {H.}~\bibnamefont {Peng}}, \bibinfo {author} {\bibfnamefont {J.}~\bibnamefont {Guo}},\ and\ \bibinfo {author} {\bibfnamefont {X.}~\bibnamefont {Zhu}},\ }\bibfield  {title} {\bibinfo {title} {Observation of the nonanalytic behavior of optical phonons in monolayer hexagonal boron nitride},\ }\href {https://www.nature.com/articles/s41467-024-46229-4} {\bibfield  {journal} {\bibinfo  {journal} {Nat. Commun.}\ }\textbf {\bibinfo
  {volume} {15}} (\bibinfo {year} {2024}{\natexlab{a}})}\BibitemShut {NoStop}%
\bibitem [{\citenamefont {Sio}\ and\ \citenamefont {Giustino}(2022)}]{sio2022unified}%
  \BibitemOpen
  \bibfield  {author} {\bibinfo {author} {\bibfnamefont {W.~H.}\ \bibnamefont {Sio}}\ and\ \bibinfo {author} {\bibfnamefont {F.}~\bibnamefont {Giustino}},\ }\bibfield  {title} {\bibinfo {title} {Unified ab initio description of fr{\"o}hlich electron-phonon interactions in two-dimensional and three-dimensional materials},\ }\href {https://journals.aps.org/prb/abstract/10.1103/PhysRevB.105.115414} {\bibfield  {journal} {\bibinfo  {journal} {Physical Review B}\ }\textbf {\bibinfo {volume} {105}},\ \bibinfo {pages} {115414} (\bibinfo {year} {2022})}\BibitemShut {NoStop}%
\bibitem [{\citenamefont {Sio}\ and\ \citenamefont {Giustino}(2023)}]{sio2023polarons}%
  \BibitemOpen
  \bibfield  {author} {\bibinfo {author} {\bibfnamefont {W.~H.}\ \bibnamefont {Sio}}\ and\ \bibinfo {author} {\bibfnamefont {F.}~\bibnamefont {Giustino}},\ }\bibfield  {title} {\bibinfo {title} {Polarons in two-dimensional atomic crystals},\ }\href {https://www.nature.com/articles/s41567-023-01953-4} {\bibfield  {journal} {\bibinfo  {journal} {Nature Physics}\ }\textbf {\bibinfo {volume} {19}},\ \bibinfo {pages} {629} (\bibinfo {year} {2023})}\BibitemShut {NoStop}%
\bibitem [{\citenamefont {Kudlis}\ \emph {et~al.}(2025)\citenamefont {Kudlis}, \citenamefont {Shahnazaryan},\ and\ \citenamefont {Tokatly}}]{kudlis2025polarons}%
  \BibitemOpen
  \bibfield  {author} {\bibinfo {author} {\bibfnamefont {A.}~\bibnamefont {Kudlis}}, \bibinfo {author} {\bibfnamefont {V.}~\bibnamefont {Shahnazaryan}},\ and\ \bibinfo {author} {\bibfnamefont {I.}~\bibnamefont {Tokatly}},\ }\bibfield  {title} {\bibinfo {title} {Polarons in two-dimensional polar materials: All-coupling variational theory},\ }\href {https://arxiv.org/abs/2507.10930} {\bibfield  {journal} {\bibinfo  {journal} {arXiv preprint arXiv:2507.10930}\ } (\bibinfo {year} {2025})}\BibitemShut {NoStop}%
\bibitem [{\citenamefont {Lee}\ \emph {et~al.}(2024)\citenamefont {Lee}, \citenamefont {Alvertis}, \citenamefont {Li}, \citenamefont {Louie}, \citenamefont {Filip}, \citenamefont {Neaton},\ and\ \citenamefont {Kioupakis}}]{lee2024phonon}%
  \BibitemOpen
  \bibfield  {author} {\bibinfo {author} {\bibfnamefont {W.}~\bibnamefont {Lee}}, \bibinfo {author} {\bibfnamefont {A.~M.}\ \bibnamefont {Alvertis}}, \bibinfo {author} {\bibfnamefont {Z.}~\bibnamefont {Li}}, \bibinfo {author} {\bibfnamefont {S.~G.}\ \bibnamefont {Louie}}, \bibinfo {author} {\bibfnamefont {M.~R.}\ \bibnamefont {Filip}}, \bibinfo {author} {\bibfnamefont {J.~B.}\ \bibnamefont {Neaton}},\ and\ \bibinfo {author} {\bibfnamefont {E.}~\bibnamefont {Kioupakis}},\ }\bibfield  {title} {\bibinfo {title} {Phonon screening of excitons in atomically thin semiconductors},\ }\href {https://journals.aps.org/prl/abstract/10.1103/PhysRevLett.133.206901} {\bibfield  {journal} {\bibinfo  {journal} {Physical Review Letters}\ }\textbf {\bibinfo {volume} {133}},\ \bibinfo {pages} {206901} (\bibinfo {year} {2024})}\BibitemShut {NoStop}%
\bibitem [{\citenamefont {Alexandrov}\ and\ \citenamefont {Ranninger}(1981)}]{alexandrov1981theory}%
  \BibitemOpen
  \bibfield  {author} {\bibinfo {author} {\bibfnamefont {A.}~\bibnamefont {Alexandrov}}\ and\ \bibinfo {author} {\bibfnamefont {J.}~\bibnamefont {Ranninger}},\ }\bibfield  {title} {\bibinfo {title} {Theory of bipolarons and bipolaronic bands},\ }\href {https://journals.aps.org/prb/abstract/10.1103/PhysRevB.23.1796} {\bibfield  {journal} {\bibinfo  {journal} {Physical Review B}\ }\textbf {\bibinfo {volume} {23}},\ \bibinfo {pages} {1796} (\bibinfo {year} {1981})}\BibitemShut {NoStop}%
\bibitem [{\citenamefont {Adamowski}(1989)}]{adamowski1989formation}%
  \BibitemOpen
  \bibfield  {author} {\bibinfo {author} {\bibfnamefont {J.}~\bibnamefont {Adamowski}},\ }\bibfield  {title} {\bibinfo {title} {Formation of fr{\"o}hlich bipolarons},\ }\href {https://journals.aps.org/prb/abstract/10.1103/PhysRevB.39.3649} {\bibfield  {journal} {\bibinfo  {journal} {Physical Review B}\ }\textbf {\bibinfo {volume} {39}},\ \bibinfo {pages} {3649} (\bibinfo {year} {1989})}\BibitemShut {NoStop}%
\bibitem [{\citenamefont {Fomin}\ and\ \citenamefont {Smondyrev}(1994)}]{fomin1994bipolaron}%
  \BibitemOpen
  \bibfield  {author} {\bibinfo {author} {\bibfnamefont {V.~M.}\ \bibnamefont {Fomin}}\ and\ \bibinfo {author} {\bibfnamefont {M.~A.}\ \bibnamefont {Smondyrev}},\ }\bibfield  {title} {\bibinfo {title} {Bipolaron confinement in two-dimensional layers},\ }\href {https://journals.aps.org/prb/abstract/10.1103/PhysRevB.49.12748} {\bibfield  {journal} {\bibinfo  {journal} {Physical Review B}\ }\textbf {\bibinfo {volume} {49}},\ \bibinfo {pages} {12748} (\bibinfo {year} {1994})}\BibitemShut {NoStop}%
\bibitem [{\citenamefont {Macridin}\ \emph {et~al.}(2004)\citenamefont {Macridin}, \citenamefont {Sawatzky},\ and\ \citenamefont {Jarrell}}]{macridin2004two}%
  \BibitemOpen
  \bibfield  {author} {\bibinfo {author} {\bibfnamefont {A.}~\bibnamefont {Macridin}}, \bibinfo {author} {\bibfnamefont {G.}~\bibnamefont {Sawatzky}},\ and\ \bibinfo {author} {\bibfnamefont {M.}~\bibnamefont {Jarrell}},\ }\bibfield  {title} {\bibinfo {title} {Two-dimensional hubbard-holstein bipolaron},\ }\href {https://journals.aps.org/prb/abstract/10.1103/PhysRevB.69.245111} {\bibfield  {journal} {\bibinfo  {journal} {Physical Review B}\ }\textbf {\bibinfo {volume} {69}},\ \bibinfo {pages} {245111} (\bibinfo {year} {2004})}\BibitemShut {NoStop}%
\bibitem [{\citenamefont {Zhang}\ \emph {et~al.}(2023)\citenamefont {Zhang}, \citenamefont {Sous}, \citenamefont {Reichman}, \citenamefont {Berciu}, \citenamefont {Millis}, \citenamefont {Prokof’ev},\ and\ \citenamefont {Svistunov}}]{zhang2023bipolaronic}%
  \BibitemOpen
  \bibfield  {author} {\bibinfo {author} {\bibfnamefont {C.}~\bibnamefont {Zhang}}, \bibinfo {author} {\bibfnamefont {J.}~\bibnamefont {Sous}}, \bibinfo {author} {\bibfnamefont {D.}~\bibnamefont {Reichman}}, \bibinfo {author} {\bibfnamefont {M.}~\bibnamefont {Berciu}}, \bibinfo {author} {\bibfnamefont {A.}~\bibnamefont {Millis}}, \bibinfo {author} {\bibfnamefont {N.}~\bibnamefont {Prokof’ev}},\ and\ \bibinfo {author} {\bibfnamefont {B.}~\bibnamefont {Svistunov}},\ }\bibfield  {title} {\bibinfo {title} {Bipolaronic high-temperature superconductivity},\ }\href {https://journals.aps.org/prx/abstract/10.1103/PhysRevX.13.011010} {\bibfield  {journal} {\bibinfo  {journal} {Physical Review X}\ }\textbf {\bibinfo {volume} {13}},\ \bibinfo {pages} {011010} (\bibinfo {year} {2023})}\BibitemShut {NoStop}%
\bibitem [{\citenamefont {Mak}\ \emph {et~al.}(2013)\citenamefont {Mak}, \citenamefont {He}, \citenamefont {Lee}, \citenamefont {Lee}, \citenamefont {Hone}, \citenamefont {Heinz},\ and\ \citenamefont {Shan}}]{mak2013tightly}%
  \BibitemOpen
  \bibfield  {author} {\bibinfo {author} {\bibfnamefont {K.~F.}\ \bibnamefont {Mak}}, \bibinfo {author} {\bibfnamefont {K.}~\bibnamefont {He}}, \bibinfo {author} {\bibfnamefont {C.}~\bibnamefont {Lee}}, \bibinfo {author} {\bibfnamefont {G.~H.}\ \bibnamefont {Lee}}, \bibinfo {author} {\bibfnamefont {J.}~\bibnamefont {Hone}}, \bibinfo {author} {\bibfnamefont {T.~F.}\ \bibnamefont {Heinz}},\ and\ \bibinfo {author} {\bibfnamefont {J.}~\bibnamefont {Shan}},\ }\bibfield  {title} {\bibinfo {title} {Tightly bound trions in monolayer mos2},\ }\href {https://www.nature.com/articles/nmat3505} {\bibfield  {journal} {\bibinfo  {journal} {Nature materials}\ }\textbf {\bibinfo {volume} {12}},\ \bibinfo {pages} {207} (\bibinfo {year} {2013})}\BibitemShut {NoStop}%
\bibitem [{\citenamefont {Wang}\ \emph {et~al.}(2018)\citenamefont {Wang}, \citenamefont {Chernikov}, \citenamefont {Glazov}, \citenamefont {Heinz}, \citenamefont {Marie}, \citenamefont {Amand},\ and\ \citenamefont {Urbaszek}}]{wang2018colloquium}%
  \BibitemOpen
  \bibfield  {author} {\bibinfo {author} {\bibfnamefont {G.}~\bibnamefont {Wang}}, \bibinfo {author} {\bibfnamefont {A.}~\bibnamefont {Chernikov}}, \bibinfo {author} {\bibfnamefont {M.~M.}\ \bibnamefont {Glazov}}, \bibinfo {author} {\bibfnamefont {T.~F.}\ \bibnamefont {Heinz}}, \bibinfo {author} {\bibfnamefont {X.}~\bibnamefont {Marie}}, \bibinfo {author} {\bibfnamefont {T.}~\bibnamefont {Amand}},\ and\ \bibinfo {author} {\bibfnamefont {B.}~\bibnamefont {Urbaszek}},\ }\bibfield  {title} {\bibinfo {title} {Colloquium: Excitons in atomically thin transition metal dichalcogenides},\ }\href {https://journals.aps.org/rmp/abstract/10.1103/RevModPhys.90.021001} {\bibfield  {journal} {\bibinfo  {journal} {Reviews of Modern Physics}\ }\textbf {\bibinfo {volume} {90}},\ \bibinfo {pages} {021001} (\bibinfo {year} {2018})}\BibitemShut {NoStop}%
\bibitem [{\citenamefont {Ziegler}\ \emph {et~al.}(2023)\citenamefont {Ziegler}, \citenamefont {Cho}, \citenamefont {Terres}, \citenamefont {Menahem}, \citenamefont {Taniguchi}, \citenamefont {Watanabe}, \citenamefont {Yaffe}, \citenamefont {Berkelbach},\ and\ \citenamefont {Chernikov}}]{ziegler2023mobile}%
  \BibitemOpen
  \bibfield  {author} {\bibinfo {author} {\bibfnamefont {J.~D.}\ \bibnamefont {Ziegler}}, \bibinfo {author} {\bibfnamefont {Y.}~\bibnamefont {Cho}}, \bibinfo {author} {\bibfnamefont {S.}~\bibnamefont {Terres}}, \bibinfo {author} {\bibfnamefont {M.}~\bibnamefont {Menahem}}, \bibinfo {author} {\bibfnamefont {T.}~\bibnamefont {Taniguchi}}, \bibinfo {author} {\bibfnamefont {K.}~\bibnamefont {Watanabe}}, \bibinfo {author} {\bibfnamefont {O.}~\bibnamefont {Yaffe}}, \bibinfo {author} {\bibfnamefont {T.~C.}\ \bibnamefont {Berkelbach}},\ and\ \bibinfo {author} {\bibfnamefont {A.}~\bibnamefont {Chernikov}},\ }\bibfield  {title} {\bibinfo {title} {Mobile trions in electrically tunable 2d hybrid perovskites},\ }\href {https://advanced.onlinelibrary.wiley.com/doi/full/10.1002/adma.202210221} {\bibfield  {journal} {\bibinfo  {journal} {Advanced Materials}\ }\textbf {\bibinfo {volume} {35}},\ \bibinfo {pages} {2210221} (\bibinfo {year} {2023})}\BibitemShut {NoStop}%
\bibitem [{\citenamefont {Makarov}\ \emph {et~al.}(2016)\citenamefont {Makarov}, \citenamefont {Guo}, \citenamefont {Isaienko}, \citenamefont {Liu}, \citenamefont {Robel},\ and\ \citenamefont {Klimov}}]{makarov2016spectral}%
  \BibitemOpen
  \bibfield  {author} {\bibinfo {author} {\bibfnamefont {N.~S.}\ \bibnamefont {Makarov}}, \bibinfo {author} {\bibfnamefont {S.}~\bibnamefont {Guo}}, \bibinfo {author} {\bibfnamefont {O.}~\bibnamefont {Isaienko}}, \bibinfo {author} {\bibfnamefont {W.}~\bibnamefont {Liu}}, \bibinfo {author} {\bibfnamefont {I.}~\bibnamefont {Robel}},\ and\ \bibinfo {author} {\bibfnamefont {V.~I.}\ \bibnamefont {Klimov}},\ }\bibfield  {title} {\bibinfo {title} {Spectral and dynamical properties of single excitons, biexcitons, and trions in cesium--lead-halide perovskite quantum dots},\ }\href {https://pubs.acs.org/doi/10.1021/acs.nanolett.5b05077} {\bibfield  {journal} {\bibinfo  {journal} {Nano letters}\ }\textbf {\bibinfo {volume} {16}},\ \bibinfo {pages} {2349} (\bibinfo {year} {2016})}\BibitemShut {NoStop}%
\bibitem [{\citenamefont {Fu}\ \emph {et~al.}(2017)\citenamefont {Fu}, \citenamefont {Tamarat}, \citenamefont {Huang}, \citenamefont {Even}, \citenamefont {Rogach},\ and\ \citenamefont {Lounis}}]{fu2017neutral}%
  \BibitemOpen
  \bibfield  {author} {\bibinfo {author} {\bibfnamefont {M.}~\bibnamefont {Fu}}, \bibinfo {author} {\bibfnamefont {P.}~\bibnamefont {Tamarat}}, \bibinfo {author} {\bibfnamefont {H.}~\bibnamefont {Huang}}, \bibinfo {author} {\bibfnamefont {J.}~\bibnamefont {Even}}, \bibinfo {author} {\bibfnamefont {A.~L.}\ \bibnamefont {Rogach}},\ and\ \bibinfo {author} {\bibfnamefont {B.}~\bibnamefont {Lounis}},\ }\bibfield  {title} {\bibinfo {title} {Neutral and charged exciton fine structure in single lead halide perovskite nanocrystals revealed by magneto-optical spectroscopy},\ }\href {https://pubs.acs.org/doi/10.1021/acs.nanolett.7b00064} {\bibfield  {journal} {\bibinfo  {journal} {Nano letters}\ }\textbf {\bibinfo {volume} {17}},\ \bibinfo {pages} {2895} (\bibinfo {year} {2017})}\BibitemShut {NoStop}%
\bibitem [{\citenamefont {Becker}\ \emph {et~al.}(2018)\citenamefont {Becker}, \citenamefont {Vaxenburg}, \citenamefont {Nedelcu}, \citenamefont {Sercel}, \citenamefont {Shabaev}, \citenamefont {Mehl}, \citenamefont {Michopoulos}, \citenamefont {Lambrakos}, \citenamefont {Bernstein}, \citenamefont {Lyons} \emph {et~al.}}]{becker2018bright}%
  \BibitemOpen
  \bibfield  {author} {\bibinfo {author} {\bibfnamefont {M.~A.}\ \bibnamefont {Becker}}, \bibinfo {author} {\bibfnamefont {R.}~\bibnamefont {Vaxenburg}}, \bibinfo {author} {\bibfnamefont {G.}~\bibnamefont {Nedelcu}}, \bibinfo {author} {\bibfnamefont {P.~C.}\ \bibnamefont {Sercel}}, \bibinfo {author} {\bibfnamefont {A.}~\bibnamefont {Shabaev}}, \bibinfo {author} {\bibfnamefont {M.~J.}\ \bibnamefont {Mehl}}, \bibinfo {author} {\bibfnamefont {J.~G.}\ \bibnamefont {Michopoulos}}, \bibinfo {author} {\bibfnamefont {S.~G.}\ \bibnamefont {Lambrakos}}, \bibinfo {author} {\bibfnamefont {N.}~\bibnamefont {Bernstein}}, \bibinfo {author} {\bibfnamefont {J.~L.}\ \bibnamefont {Lyons}}, \emph {et~al.},\ }\bibfield  {title} {\bibinfo {title} {Bright triplet excitons in caesium lead halide perovskites},\ }\href {https://www.nature.com/articles/nature25147} {\bibfield  {journal} {\bibinfo  {journal} {Nature}\ }\textbf {\bibinfo {volume} {553}},\ \bibinfo {pages} {189} (\bibinfo {year} {2018})}\BibitemShut {NoStop}%
\bibitem [{\citenamefont {Zhu}\ \emph {et~al.}(2023)\citenamefont {Zhu}, \citenamefont {Nguyen}, \citenamefont {Boehme}, \citenamefont {Moskalenko}, \citenamefont {Dirin}, \citenamefont {Bodnarchuk}, \citenamefont {Katan}, \citenamefont {Even}, \citenamefont {Rain{\`o}},\ and\ \citenamefont {Kovalenko}}]{zhu2023many}%
  \BibitemOpen
  \bibfield  {author} {\bibinfo {author} {\bibfnamefont {C.}~\bibnamefont {Zhu}}, \bibinfo {author} {\bibfnamefont {T.}~\bibnamefont {Nguyen}}, \bibinfo {author} {\bibfnamefont {S.~C.}\ \bibnamefont {Boehme}}, \bibinfo {author} {\bibfnamefont {A.}~\bibnamefont {Moskalenko}}, \bibinfo {author} {\bibfnamefont {D.~N.}\ \bibnamefont {Dirin}}, \bibinfo {author} {\bibfnamefont {M.~I.}\ \bibnamefont {Bodnarchuk}}, \bibinfo {author} {\bibfnamefont {C.}~\bibnamefont {Katan}}, \bibinfo {author} {\bibfnamefont {J.}~\bibnamefont {Even}}, \bibinfo {author} {\bibfnamefont {G.}~\bibnamefont {Rain{\`o}}},\ and\ \bibinfo {author} {\bibfnamefont {M.~V.}\ \bibnamefont {Kovalenko}},\ }\bibfield  {title} {\bibinfo {title} {Many-body correlations and exciton complexes in cspbbr 3 quantum dots},\ }\href {https://advanced.onlinelibrary.wiley.com/doi/full/10.1002/adma.202208354} {\bibfield  {journal} {\bibinfo  {journal} {Advanced Materials}\ }\textbf {\bibinfo {volume} {35}},\ \bibinfo {pages} {2208354} (\bibinfo {year}
  {2023})}\BibitemShut {NoStop}%
\bibitem [{\citenamefont {Tamarat}\ \emph {et~al.}(2023)\citenamefont {Tamarat}, \citenamefont {Prin}, \citenamefont {Berezovska}, \citenamefont {Moskalenko}, \citenamefont {Nguyen}, \citenamefont {Xia}, \citenamefont {Hou}, \citenamefont {Trebbia}, \citenamefont {Zacharias}, \citenamefont {Pedesseau} \emph {et~al.}}]{tamarat2023universal}%
  \BibitemOpen
  \bibfield  {author} {\bibinfo {author} {\bibfnamefont {P.}~\bibnamefont {Tamarat}}, \bibinfo {author} {\bibfnamefont {E.}~\bibnamefont {Prin}}, \bibinfo {author} {\bibfnamefont {Y.}~\bibnamefont {Berezovska}}, \bibinfo {author} {\bibfnamefont {A.}~\bibnamefont {Moskalenko}}, \bibinfo {author} {\bibfnamefont {T.~P.~T.}\ \bibnamefont {Nguyen}}, \bibinfo {author} {\bibfnamefont {C.}~\bibnamefont {Xia}}, \bibinfo {author} {\bibfnamefont {L.}~\bibnamefont {Hou}}, \bibinfo {author} {\bibfnamefont {J.-B.}\ \bibnamefont {Trebbia}}, \bibinfo {author} {\bibfnamefont {M.}~\bibnamefont {Zacharias}}, \bibinfo {author} {\bibfnamefont {L.}~\bibnamefont {Pedesseau}}, \emph {et~al.},\ }\bibfield  {title} {\bibinfo {title} {Universal scaling laws for charge-carrier interactions with quantum confinement in lead-halide perovskites},\ }\href {https://www.nature.com/articles/s41467-023-35842-4} {\bibfield  {journal} {\bibinfo  {journal} {Nature Communications}\ }\textbf {\bibinfo {volume} {14}},\ \bibinfo {pages} {229} (\bibinfo
  {year} {2023})}\BibitemShut {NoStop}%
\bibitem [{\citenamefont {Cho}\ \emph {et~al.}(2024)\citenamefont {Cho}, \citenamefont {Sato}, \citenamefont {Yamada}, \citenamefont {Sato}, \citenamefont {Saruyama}, \citenamefont {Teranishi}, \citenamefont {Suzuura},\ and\ \citenamefont {Kanemitsu}}]{cho2024size}%
  \BibitemOpen
  \bibfield  {author} {\bibinfo {author} {\bibfnamefont {K.}~\bibnamefont {Cho}}, \bibinfo {author} {\bibfnamefont {T.}~\bibnamefont {Sato}}, \bibinfo {author} {\bibfnamefont {T.}~\bibnamefont {Yamada}}, \bibinfo {author} {\bibfnamefont {R.}~\bibnamefont {Sato}}, \bibinfo {author} {\bibfnamefont {M.}~\bibnamefont {Saruyama}}, \bibinfo {author} {\bibfnamefont {T.}~\bibnamefont {Teranishi}}, \bibinfo {author} {\bibfnamefont {H.}~\bibnamefont {Suzuura}},\ and\ \bibinfo {author} {\bibfnamefont {Y.}~\bibnamefont {Kanemitsu}},\ }\bibfield  {title} {\bibinfo {title} {Size dependence of trion and biexciton binding energies in lead halide perovskite nanocrystals},\ }\href {https://pubs.acs.org/doi/abs/10.1021/acsnano.3c11842} {\bibfield  {journal} {\bibinfo  {journal} {ACS nano}\ }\textbf {\bibinfo {volume} {18}},\ \bibinfo {pages} {5723} (\bibinfo {year} {2024})}\BibitemShut {NoStop}%
\bibitem [{\citenamefont {Movilla}\ \emph {et~al.}(2025)\citenamefont {Movilla}, \citenamefont {Planelles},\ and\ \citenamefont {Climente}}]{movilla2025binding}%
  \BibitemOpen
  \bibfield  {author} {\bibinfo {author} {\bibfnamefont {J.~L.}\ \bibnamefont {Movilla}}, \bibinfo {author} {\bibfnamefont {J.}~\bibnamefont {Planelles}},\ and\ \bibinfo {author} {\bibfnamefont {J.~I.}\ \bibnamefont {Climente}},\ }\bibfield  {title} {\bibinfo {title} {Binding energy of polaronic trions and biexcitons in cspbbr 3 nanocrystals},\ }\href {https://journals.aps.org/prb/abstract/10.1103/PhysRevB.111.155304} {\bibfield  {journal} {\bibinfo  {journal} {Physical Review B}\ }\textbf {\bibinfo {volume} {111}},\ \bibinfo {pages} {155304} (\bibinfo {year} {2025})}\BibitemShut {NoStop}%
\bibitem [{\citenamefont {Kittel}(1987)}]{kittel1987quantum}%
  \BibitemOpen
  \bibfield  {author} {\bibinfo {author} {\bibfnamefont {C.}~\bibnamefont {Kittel}},\ }\href {https://www.wiley.com/en-us/Quantum+Theory+of+Solids%2C+2nd+Revised+Edition-p-9780471624127} {\emph {\bibinfo {title} {Quantum Theory of Solids}}}\ (\bibinfo  {publisher} {Wiley},\ \bibinfo {year} {1987})\BibitemShut {NoStop}%
\bibitem [{\citenamefont {Soufiani}\ \emph {et~al.}(2015)\citenamefont {Soufiani}, \citenamefont {Huang}, \citenamefont {Reece}, \citenamefont {Sheng}, \citenamefont {Ho-Baillie},\ and\ \citenamefont {Green}}]{soufiani2015polaronic}%
  \BibitemOpen
  \bibfield  {author} {\bibinfo {author} {\bibfnamefont {A.~M.}\ \bibnamefont {Soufiani}}, \bibinfo {author} {\bibfnamefont {F.}~\bibnamefont {Huang}}, \bibinfo {author} {\bibfnamefont {P.}~\bibnamefont {Reece}}, \bibinfo {author} {\bibfnamefont {R.}~\bibnamefont {Sheng}}, \bibinfo {author} {\bibfnamefont {A.}~\bibnamefont {Ho-Baillie}},\ and\ \bibinfo {author} {\bibfnamefont {M.~A.}\ \bibnamefont {Green}},\ }\bibfield  {title} {\bibinfo {title} {Polaronic exciton binding energy in iodide and bromide organic-inorganic lead halide perovskites},\ }\href {https://pubs.aip.org/aip/apl/article-abstract/107/23/231902/30336/Polaronic-exciton-binding-energy-in-iodide-and} {\bibfield  {journal} {\bibinfo  {journal} {Applied Physics Letters}\ }\textbf {\bibinfo {volume} {107}} (\bibinfo {year} {2015})}\BibitemShut {NoStop}%
\bibitem [{\citenamefont {Bokdam}\ \emph {et~al.}(2016)\citenamefont {Bokdam}, \citenamefont {Sander}, \citenamefont {Stroppa}, \citenamefont {Picozzi}, \citenamefont {Sarma}, \citenamefont {Franchini},\ and\ \citenamefont {Kresse}}]{bokdam2016role}%
  \BibitemOpen
  \bibfield  {author} {\bibinfo {author} {\bibfnamefont {M.}~\bibnamefont {Bokdam}}, \bibinfo {author} {\bibfnamefont {T.}~\bibnamefont {Sander}}, \bibinfo {author} {\bibfnamefont {A.}~\bibnamefont {Stroppa}}, \bibinfo {author} {\bibfnamefont {S.}~\bibnamefont {Picozzi}}, \bibinfo {author} {\bibfnamefont {D.}~\bibnamefont {Sarma}}, \bibinfo {author} {\bibfnamefont {C.}~\bibnamefont {Franchini}},\ and\ \bibinfo {author} {\bibfnamefont {G.}~\bibnamefont {Kresse}},\ }\bibfield  {title} {\bibinfo {title} {Role of polar phonons in the photo excited state of metal halide perovskites},\ }\href {https://www.nature.com/articles/srep28618} {\bibfield  {journal} {\bibinfo  {journal} {Scientific reports}\ }\textbf {\bibinfo {volume} {6}},\ \bibinfo {pages} {28618} (\bibinfo {year} {2016})}\BibitemShut {NoStop}%
\bibitem [{\citenamefont {Nagai}\ \emph {et~al.}(2018)\citenamefont {Nagai}, \citenamefont {Tomioka}, \citenamefont {Ashida}, \citenamefont {Hoyano}, \citenamefont {Akashi}, \citenamefont {Yamada}, \citenamefont {Aharen},\ and\ \citenamefont {Kanemitsu}}]{nagai2018longitudinal}%
  \BibitemOpen
  \bibfield  {author} {\bibinfo {author} {\bibfnamefont {M.}~\bibnamefont {Nagai}}, \bibinfo {author} {\bibfnamefont {T.}~\bibnamefont {Tomioka}}, \bibinfo {author} {\bibfnamefont {M.}~\bibnamefont {Ashida}}, \bibinfo {author} {\bibfnamefont {M.}~\bibnamefont {Hoyano}}, \bibinfo {author} {\bibfnamefont {R.}~\bibnamefont {Akashi}}, \bibinfo {author} {\bibfnamefont {Y.}~\bibnamefont {Yamada}}, \bibinfo {author} {\bibfnamefont {T.}~\bibnamefont {Aharen}},\ and\ \bibinfo {author} {\bibfnamefont {Y.}~\bibnamefont {Kanemitsu}},\ }\bibfield  {title} {\bibinfo {title} {Longitudinal optical phonons modified by organic molecular cation motions in organic-inorganic hybrid perovskites},\ }\href {https://journals.aps.org/prl/abstract/10.1103/PhysRevLett.121.145506} {\bibfield  {journal} {\bibinfo  {journal} {Physical review letters}\ }\textbf {\bibinfo {volume} {121}},\ \bibinfo {pages} {145506} (\bibinfo {year} {2018})}\BibitemShut {NoStop}%
\bibitem [{\citenamefont {Sendner}\ \emph {et~al.}(2016)\citenamefont {Sendner}, \citenamefont {Nayak}, \citenamefont {Egger}, \citenamefont {Beck}, \citenamefont {M{\"u}ller}, \citenamefont {Epding}, \citenamefont {Kowalsky}, \citenamefont {Kronik}, \citenamefont {Snaith}, \citenamefont {Pucci} \emph {et~al.}}]{sendner2016optical}%
  \BibitemOpen
  \bibfield  {author} {\bibinfo {author} {\bibfnamefont {M.}~\bibnamefont {Sendner}}, \bibinfo {author} {\bibfnamefont {P.~K.}\ \bibnamefont {Nayak}}, \bibinfo {author} {\bibfnamefont {D.~A.}\ \bibnamefont {Egger}}, \bibinfo {author} {\bibfnamefont {S.}~\bibnamefont {Beck}}, \bibinfo {author} {\bibfnamefont {C.}~\bibnamefont {M{\"u}ller}}, \bibinfo {author} {\bibfnamefont {B.}~\bibnamefont {Epding}}, \bibinfo {author} {\bibfnamefont {W.}~\bibnamefont {Kowalsky}}, \bibinfo {author} {\bibfnamefont {L.}~\bibnamefont {Kronik}}, \bibinfo {author} {\bibfnamefont {H.~J.}\ \bibnamefont {Snaith}}, \bibinfo {author} {\bibfnamefont {A.}~\bibnamefont {Pucci}}, \emph {et~al.},\ }\bibfield  {title} {\bibinfo {title} {Optical phonons in methylammonium lead halide perovskites and implications for charge transport},\ }\href {https://pubs.rsc.org/en/content/articlelanding/2016/mh/c6mh00275g} {\bibfield  {journal} {\bibinfo  {journal} {Materials Horizons}\ }\textbf {\bibinfo {volume} {3}},\ \bibinfo {pages} {613} (\bibinfo
  {year} {2016})}\BibitemShut {NoStop}%
\bibitem [{\citenamefont {Sanders}\ \emph {et~al.}(2017)\citenamefont {Sanders}, \citenamefont {Bayerl}, \citenamefont {Shi}, \citenamefont {Mengle},\ and\ \citenamefont {Kioupakis}}]{Sanders2017}%
  \BibitemOpen
  \bibfield  {author} {\bibinfo {author} {\bibfnamefont {N.}~\bibnamefont {Sanders}}, \bibinfo {author} {\bibfnamefont {D.}~\bibnamefont {Bayerl}}, \bibinfo {author} {\bibfnamefont {G.}~\bibnamefont {Shi}}, \bibinfo {author} {\bibfnamefont {K.~A.}\ \bibnamefont {Mengle}},\ and\ \bibinfo {author} {\bibfnamefont {E.}~\bibnamefont {Kioupakis}},\ }\bibfield  {title} {\bibinfo {title} {Electronic and optical properties of two-dimensional gan from first-principles},\ }\href {https://doi.org/10.1021/acs.nanolett.7b03003} {\bibfield  {journal} {\bibinfo  {journal} {Nano Letters}\ }\textbf {\bibinfo {volume} {17}},\ \bibinfo {pages} {7345–7349} (\bibinfo {year} {2017})}\BibitemShut {NoStop}%
\bibitem [{\citenamefont {Zhao}\ \emph {et~al.}(2016)\citenamefont {Zhao}, \citenamefont {Xu}, \citenamefont {Wang},\ and\ \citenamefont {Lin}}]{doi:10.1021/acs.jpcc.6b09706}%
  \BibitemOpen
  \bibfield  {author} {\bibinfo {author} {\bibfnamefont {L.}~\bibnamefont {Zhao}}, \bibinfo {author} {\bibfnamefont {S.}~\bibnamefont {Xu}}, \bibinfo {author} {\bibfnamefont {M.}~\bibnamefont {Wang}},\ and\ \bibinfo {author} {\bibfnamefont {S.}~\bibnamefont {Lin}},\ }\bibfield  {title} {\bibinfo {title} {Probing the thermodynamic stability and phonon transport in two-dimensional hexagonal aluminum nitride monolayer},\ }\href {https://doi.org/10.1021/acs.jpcc.6b09706} {\bibfield  {journal} {\bibinfo  {journal} {The Journal of Physical Chemistry C}\ }\textbf {\bibinfo {volume} {120}},\ \bibinfo {pages} {27675} (\bibinfo {year} {2016})},\ \Eprint {https://arxiv.org/abs/https://doi.org/10.1021/acs.jpcc.6b09706} {https://doi.org/10.1021/acs.jpcc.6b09706} \BibitemShut {NoStop}%
\bibitem [{\citenamefont {Li}\ \emph {et~al.}(2024{\natexlab{b}})\citenamefont {Li}, \citenamefont {Xue}, \citenamefont {Zhao},\ and\ \citenamefont {Lu}}]{Li_2024}%
  \BibitemOpen
  \bibfield  {author} {\bibinfo {author} {\bibfnamefont {W.}~\bibnamefont {Li}}, \bibinfo {author} {\bibfnamefont {F.-n.}\ \bibnamefont {Xue}}, \bibinfo {author} {\bibfnamefont {P.-b.}\ \bibnamefont {Zhao}},\ and\ \bibinfo {author} {\bibfnamefont {Y.}~\bibnamefont {Lu}},\ }\bibfield  {title} {\bibinfo {title} {Electronic structure and phonon transport properties of hfse2 under in-plane strain and finite temperature},\ }\href {https://doi.org/10.1088/1402-4896/ad720b} {\bibfield  {journal} {\bibinfo  {journal} {Physica Scripta}\ }\textbf {\bibinfo {volume} {99}},\ \bibinfo {pages} {105916} (\bibinfo {year} {2024}{\natexlab{b}})}\BibitemShut {NoStop}%
\bibitem [{\citenamefont {Wang}\ \emph {et~al.}(2021)\citenamefont {Wang}, \citenamefont {Lan}, \citenamefont {Dai}, \citenamefont {Zhang}, \citenamefont {Wang},\ and\ \citenamefont {Ge}}]{doi:10.1021/acsomega.1c04286}%
  \BibitemOpen
  \bibfield  {author} {\bibinfo {author} {\bibfnamefont {H.}~\bibnamefont {Wang}}, \bibinfo {author} {\bibfnamefont {Y.-S.}\ \bibnamefont {Lan}}, \bibinfo {author} {\bibfnamefont {B.}~\bibnamefont {Dai}}, \bibinfo {author} {\bibfnamefont {X.-W.}\ \bibnamefont {Zhang}}, \bibinfo {author} {\bibfnamefont {Z.-G.}\ \bibnamefont {Wang}},\ and\ \bibinfo {author} {\bibfnamefont {N.-N.}\ \bibnamefont {Ge}},\ }\bibfield  {title} {\bibinfo {title} {Improved thermoelectric performance of monolayer hfs2 by strain engineering},\ }\href {https://doi.org/10.1021/acsomega.1c04286} {\bibfield  {journal} {\bibinfo  {journal} {ACS Omega}\ }\textbf {\bibinfo {volume} {6}},\ \bibinfo {pages} {29820} (\bibinfo {year} {2021})},\ \bibinfo {note} {pMID: 34778655},\ \Eprint {https://arxiv.org/abs/https://doi.org/10.1021/acsomega.1c04286} {https://doi.org/10.1021/acsomega.1c04286} \BibitemShut {NoStop}%
\bibitem [{\citenamefont {Pandit}\ and\ \citenamefont {Hamad}(2021)}]{Pandit_2021}%
  \BibitemOpen
  \bibfield  {author} {\bibinfo {author} {\bibfnamefont {A.}~\bibnamefont {Pandit}}\ and\ \bibinfo {author} {\bibfnamefont {B.}~\bibnamefont {Hamad}},\ }\bibfield  {title} {\bibinfo {title} {The effect of finite-temperature and anharmonic lattice dynamics on the thermal conductivity of zrs2 monolayer: self-consistent phonon calculations},\ }\href {https://doi.org/10.1088/1361-648X/ac1822} {\bibfield  {journal} {\bibinfo  {journal} {Journal of Physics: Condensed Matter}\ }\textbf {\bibinfo {volume} {33}},\ \bibinfo {pages} {425405} (\bibinfo {year} {2021})}\BibitemShut {NoStop}%
\bibitem [{\citenamefont {Suzuki}\ and\ \citenamefont {Varga}(1998)}]{SuzukiVarga1998}%
  \BibitemOpen
  \bibfield  {author} {\bibinfo {author} {\bibfnamefont {Y.}~\bibnamefont {Suzuki}}\ and\ \bibinfo {author} {\bibfnamefont {K.}~\bibnamefont {Varga}},\ }\href@noop {} {\emph {\bibinfo {title} {Stochastic Variational Approach to Quantum-Mechanical Few-Body Problems}}},\ Springer Series on Atomic, Optical, and Plasma Physics\ (\bibinfo  {publisher} {Springer},\ \bibinfo {address} {Berlin\,/\,Heidelberg},\ \bibinfo {year} {1998})\ pp.\ \bibinfo {pages} {i--xiv, 1--314}\BibitemShut {NoStop}%
\bibitem [{\citenamefont {Courtade}\ \emph {et~al.}(2017)\citenamefont {Courtade}, \citenamefont {Semina}, \citenamefont {Manca}, \citenamefont {Glazov}, \citenamefont {Robert}, \citenamefont {Cadiz}, \citenamefont {Wang}, \citenamefont {Taniguchi}, \citenamefont {Watanabe}, \citenamefont {Pierre} \emph {et~al.}}]{courtade2017charged}%
  \BibitemOpen
  \bibfield  {author} {\bibinfo {author} {\bibfnamefont {E.}~\bibnamefont {Courtade}}, \bibinfo {author} {\bibfnamefont {M.}~\bibnamefont {Semina}}, \bibinfo {author} {\bibfnamefont {M.}~\bibnamefont {Manca}}, \bibinfo {author} {\bibfnamefont {M.}~\bibnamefont {Glazov}}, \bibinfo {author} {\bibfnamefont {C.}~\bibnamefont {Robert}}, \bibinfo {author} {\bibfnamefont {F.}~\bibnamefont {Cadiz}}, \bibinfo {author} {\bibfnamefont {G.}~\bibnamefont {Wang}}, \bibinfo {author} {\bibfnamefont {T.}~\bibnamefont {Taniguchi}}, \bibinfo {author} {\bibfnamefont {K.}~\bibnamefont {Watanabe}}, \bibinfo {author} {\bibfnamefont {M.}~\bibnamefont {Pierre}}, \emph {et~al.},\ }\bibfield  {title} {\bibinfo {title} {Charged excitons in monolayer wse 2: Experiment and theory},\ }\href {https://journals.aps.org/prb/abstract/10.1103/PhysRevB.96.085302} {\bibfield  {journal} {\bibinfo  {journal} {Physical Review B}\ }\textbf {\bibinfo {volume} {96}},\ \bibinfo {pages} {085302} (\bibinfo {year} {2017})}\BibitemShut {NoStop}%
\bibitem [{\citenamefont {Shi}\ \emph {et~al.}(2020)\citenamefont {Shi}, \citenamefont {Li}, \citenamefont {Wu}, \citenamefont {Wu}, \citenamefont {Luo}, \citenamefont {Li}, \citenamefont {Jasieniak},\ and\ \citenamefont {Meng}}]{shi2020exciton}%
  \BibitemOpen
  \bibfield  {author} {\bibinfo {author} {\bibfnamefont {J.}~\bibnamefont {Shi}}, \bibinfo {author} {\bibfnamefont {Y.}~\bibnamefont {Li}}, \bibinfo {author} {\bibfnamefont {J.}~\bibnamefont {Wu}}, \bibinfo {author} {\bibfnamefont {H.}~\bibnamefont {Wu}}, \bibinfo {author} {\bibfnamefont {Y.}~\bibnamefont {Luo}}, \bibinfo {author} {\bibfnamefont {D.}~\bibnamefont {Li}}, \bibinfo {author} {\bibfnamefont {J.~J.}\ \bibnamefont {Jasieniak}},\ and\ \bibinfo {author} {\bibfnamefont {Q.}~\bibnamefont {Meng}},\ }\bibfield  {title} {\bibinfo {title} {Exciton character and high-performance stimulated emission of hybrid lead bromide perovskite polycrystalline film},\ }\href {https://advanced.onlinelibrary.wiley.com/doi/full/10.1002/adom.201902026} {\bibfield  {journal} {\bibinfo  {journal} {Advanced Optical Materials}\ }\textbf {\bibinfo {volume} {8}},\ \bibinfo {pages} {1902026} (\bibinfo {year} {2020})}\BibitemShut {NoStop}%
\bibitem [{\citenamefont {Chernikov}\ \emph {et~al.}(2014)\citenamefont {Chernikov}, \citenamefont {Berkelbach}, \citenamefont {Hill}, \citenamefont {Rigosi}, \citenamefont {Li}, \citenamefont {Aslan}, \citenamefont {Reichman}, \citenamefont {Hybertsen},\ and\ \citenamefont {Heinz}}]{chernikov2014exciton}%
  \BibitemOpen
  \bibfield  {author} {\bibinfo {author} {\bibfnamefont {A.}~\bibnamefont {Chernikov}}, \bibinfo {author} {\bibfnamefont {T.~C.}\ \bibnamefont {Berkelbach}}, \bibinfo {author} {\bibfnamefont {H.~M.}\ \bibnamefont {Hill}}, \bibinfo {author} {\bibfnamefont {A.}~\bibnamefont {Rigosi}}, \bibinfo {author} {\bibfnamefont {Y.}~\bibnamefont {Li}}, \bibinfo {author} {\bibfnamefont {B.}~\bibnamefont {Aslan}}, \bibinfo {author} {\bibfnamefont {D.~R.}\ \bibnamefont {Reichman}}, \bibinfo {author} {\bibfnamefont {M.~S.}\ \bibnamefont {Hybertsen}},\ and\ \bibinfo {author} {\bibfnamefont {T.~F.}\ \bibnamefont {Heinz}},\ }\bibfield  {title} {\bibinfo {title} {Exciton binding energy and nonhydrogenic rydberg series in monolayer ws 2},\ }\href {https://journals.aps.org/prl/abstract/10.1103/PhysRevLett.113.076802} {\bibfield  {journal} {\bibinfo  {journal} {Physical review letters}\ }\textbf {\bibinfo {volume} {113}},\ \bibinfo {pages} {076802} (\bibinfo {year} {2014})}\BibitemShut {NoStop}%
\bibitem [{\citenamefont {Xie}\ \emph {et~al.}(2017)\citenamefont {Xie}, \citenamefont {Liao}, \citenamefont {Gong}, \citenamefont {Li}, \citenamefont {Shi}, \citenamefont {Pei}, \citenamefont {Sun}, \citenamefont {Zhang}, \citenamefont {Kong}, \citenamefont {Subbaraman} \emph {et~al.}}]{xie2017stitching}%
  \BibitemOpen
  \bibfield  {author} {\bibinfo {author} {\bibfnamefont {J.}~\bibnamefont {Xie}}, \bibinfo {author} {\bibfnamefont {L.}~\bibnamefont {Liao}}, \bibinfo {author} {\bibfnamefont {Y.}~\bibnamefont {Gong}}, \bibinfo {author} {\bibfnamefont {Y.}~\bibnamefont {Li}}, \bibinfo {author} {\bibfnamefont {F.}~\bibnamefont {Shi}}, \bibinfo {author} {\bibfnamefont {A.}~\bibnamefont {Pei}}, \bibinfo {author} {\bibfnamefont {J.}~\bibnamefont {Sun}}, \bibinfo {author} {\bibfnamefont {R.}~\bibnamefont {Zhang}}, \bibinfo {author} {\bibfnamefont {B.}~\bibnamefont {Kong}}, \bibinfo {author} {\bibfnamefont {R.}~\bibnamefont {Subbaraman}}, \emph {et~al.},\ }\bibfield  {title} {\bibinfo {title} {Stitching h-bn by atomic layer deposition of lif as a stable interface for lithium metal anode},\ }\href {https://www.science.org/doi/10.1126/sciadv.aao3170} {\bibfield  {journal} {\bibinfo  {journal} {Science advances}\ }\textbf {\bibinfo {volume} {3}},\ \bibinfo {pages} {eaao3170} (\bibinfo {year} {2017})}\BibitemShut {NoStop}%
\bibitem [{\citenamefont {Roux}\ \emph {et~al.}(2025)\citenamefont {Roux}, \citenamefont {Arnold}, \citenamefont {Carr{\'e}}, \citenamefont {Plaud}, \citenamefont {Ren}, \citenamefont {Fossard}, \citenamefont {Horezan}, \citenamefont {Janzen}, \citenamefont {Edgar}, \citenamefont {Maestre} \emph {et~al.}}]{roux2025exciton}%
  \BibitemOpen
  \bibfield  {author} {\bibinfo {author} {\bibfnamefont {S.}~\bibnamefont {Roux}}, \bibinfo {author} {\bibfnamefont {C.}~\bibnamefont {Arnold}}, \bibinfo {author} {\bibfnamefont {E.}~\bibnamefont {Carr{\'e}}}, \bibinfo {author} {\bibfnamefont {A.}~\bibnamefont {Plaud}}, \bibinfo {author} {\bibfnamefont {L.}~\bibnamefont {Ren}}, \bibinfo {author} {\bibfnamefont {F.}~\bibnamefont {Fossard}}, \bibinfo {author} {\bibfnamefont {N.}~\bibnamefont {Horezan}}, \bibinfo {author} {\bibfnamefont {E.}~\bibnamefont {Janzen}}, \bibinfo {author} {\bibfnamefont {J.~H.}\ \bibnamefont {Edgar}}, \bibinfo {author} {\bibfnamefont {C.}~\bibnamefont {Maestre}}, \emph {et~al.},\ }\bibfield  {title} {\bibinfo {title} {Exciton self-trapping in twisted hexagonal boron nitride homostructures},\ }\href {https://journals.aps.org/prx/abstract/10.1103/PhysRevX.15.021067} {\bibfield  {journal} {\bibinfo  {journal} {Physical Review X}\ }\textbf {\bibinfo {volume} {15}},\ \bibinfo {pages} {021067} (\bibinfo {year} {2025})}\BibitemShut {NoStop}%
\end{thebibliography}
\end{document}